\documentclass[12pt,reqno]{amsart}
\newtheorem{thm}{Theorem}[section]
\newtheorem{prop}[thm]{Proposition}
\newtheorem{hyp}[thm]{Hypothesis}
\newtheorem{cor}[thm]{Corollary}
\newtheorem{lem}[thm]{Lemma}

\theoremstyle{definition}
\newtheorem{defn}[thm]{Definition}

\newtheorem{exmp}[thm]{Example}
\newtheorem{rem}[thm]{Remark}

\newcommand{\bbN}{{\Bbb{N}}}

\newcommand{\bbC}{{\Bbb{C}}}
\newcommand{\calK}{{\mathcal K}}
\newcommand{\calE}{{\mathcal E}}

\newcommand{\bb}[1]{{\mathbb{#1}}}
\newcommand{\mc}[1]{{\mathcal{#1}}}

\newcommand{\be}{\begin{equation}}
\newcommand{\ee}{\end{equation}}

\newcommand{\ba}{\begin{eqnarray}}
\newcommand{\ea}{\end{eqnarray}}

\newcommand{\bma}{\begin{array}}
\newcommand{\ema}{\end{array}}
\newcommand{\lb}{\label}

\renewcommand{\Re}{\operatorname{Re}}
\renewcommand{\Im}{\operatorname{Im}}

\newcommand{\ord}{\operatorname{ord}}
\newcommand{\e}{\hbox{\rm e}}
\newcommand{\romannr}[1]{\uppercase\expandafter{\romannumeral#1}}
\newcommand{\tr}{\operatorname{tr}}
\newcommand{\diag}{\operatorname{diag}}

\makeatletter
\def\theequation{\thesection.\@arabic\c@equation}
\makeatother

\begin{document}
\title{A characterization of all elliptic solutions of the
AKNS hierarchy}
\thanks{Based upon work supported by the US National
Science Foundation under Grants No. DMS-9623121 and DMS-9401816.}
\thanks{{\it 1991 Mathematics Subject Classification.}
Primary 35Q55, 34L40;
Secondary 58F07.}
\thanks{{\it Key words and phrases.} Elliptic algebro-geometric
solutions, AKNS
hierarchy, Floquet theory, Dirac-type operators.}
\author{F.~Gesztesy${}^1$}
\address{${}^1$ Department of Mathematics,
University of Missouri,
Columbia, MO  65211, USA.}
\email{fritz@@math.missouri.edu}
\author{R.~Weikard${}^2$}
\address{${}^2$ Department of Mathematics, University of
Alabama at
Birmingham, Birmingham, AL  35294--1170, USA.}
\email{rudi@@math.uab.edu}
\date{May 14}


\begin{abstract}
An explicit characterization of all elliptic algebro-geometric solutions
of the AKNS hierarchy is presented. More precisely, we show that a pair
of elliptic functions $(p,q)$ is an algebro-geometric AKNS potential,
that is, a solution of some equation of the stationary AKNS hierarchy,
if and only if the associated linear differential system $J\Psi'
+Q\Psi=E\Psi$, where $J=\begin{pmatrix}i&0\\0&-i\end{pmatrix}$,
$Q=\begin{pmatrix}0&-iq(x)\\ip(x)&0\end{pmatrix}$, has a fundamental
system of solutions which are meromorphic with respect to the
independent variable for infinitely many and hence for all values of the
spectral parameter $E\in\bb C$.

Our approach is based on (an extension of) a classical theorem of
Picard, which  guarantees the existence of solutions which are elliptic
of the second kind for $n^{\rm th}$-order ordinary differential
equations with elliptic coefficients associated with a common period
lattice. The fundamental link between Picard's theorem and elliptic
algebro-geometric solutions of completely integrable hierarchies of
nonlinear evolution equations has recently been established in
connection with the KdV hierarchy. The current investigation appears to
be the first of its kind associated with matrix-valued Lax pairs. As
by-products we offer a detailed Floquet analysis of Dirac-type
differential expressions with periodic coefficients specifically
emphasizing algebro-geometric coefficients and a constructive reduction
of
singular
hyperelliptic curves and their Baker-Akhiezer functions to the
nonsingular case.
\end{abstract}

\maketitle

\section{Introduction} \label{intro}

Before describing our approach in some detail, we shall give a brief
account of the history of the problem of characterizing elliptic
algebro-geometric solutions of completely integrable systems. This
theme
dates back to a 1940 paper of Ince \cite{36} who studied what is
presently called the Lam\'e--Ince potential \linebreak[0]
\begin{equation} \label{1.1}
q(x)=-n(n+1)\wp(x+\omega_3), \quad n\in \bb N,
\; x \in\bb R
\end{equation}
in connection with the second-order ordinary
differential
equation
\begin{equation} \label{1.2}
y''(E,x) + q(x) y(E,x) = Ey(E,x),
\quad E\in\bb C.
\end{equation}
Here $\wp(x)=\wp(x;\omega_1,\omega_3)$ denotes the elliptic Weierstrass
function with fundamental periods $2\omega_1$ and $2\omega_3$
($\Im(\omega_3/\omega_1)\ne 0$). In the special case where $\omega_1$ is
real and $\omega_3$ is purely imaginary, the potential $q(x)$ in
(\ref{1.1}) is real-valued and Ince's striking result \cite{36}, in
modern spectral theoretic terminology, yields that the spectrum of the
unique self-adjoint operator associated with the differential expression
$L_2=d^2/dx^2 + q(x)$ in $L^2(\bb R)$ exhibits finitely many bands
(respectively gaps), that is,
\begin{equation} \label{1.3}
\sigma(L_2)=(-\infty, E_{2n}] \cup \bigcup^n_{m=1}
\left[ E_{2m-1},E_{2m-2}\right],
\quad E_{2n}<E_{2n-1}<\ldots < E_0.
\end{equation}

What we call the Lam\'e--Ince potential has, in fact, a long history and
many investigations of it precede Ince's work \cite{36}. Without
attempting to be complete we refer the interested reader, for
instance, to \cite{2b}, \cite{3}, Sect. 59, \cite{4a}, Ch. IX,
\cite{6a}, Sect. 3.6.4, \cite{13}, Sects. 135--138, \cite{13a},
\cite{13c}, \cite{13b}, \cite{29}, \cite{34}, p. 494--498, \cite{35}, p.
118--122, 266--418, 475--478, \cite{37}, p. 378--380, \cite{39a},
\cite{41}, p. 265--275, \cite{49b}, \cite{49a}, \cite{64}, \cite{66},
\cite{67}, Ch. XXIII as pertinent publications before and after Ince's
fundamental paper.

Following the traditional terminology, any real-valued potential $q$
that gives rise to a spectrum of the type (\ref{1.3}) is called an
algebro-geometric KdV potential. The proper extension of this notion to
general complex-valued meromorphic potentials $q$ then proceeds via the
KdV hierarchy of nonlinear evolution equations obtained from appropriate
Lax pairs $(P_{2n+1}(t),L_2(t))$, with $L_2(t)=d^2/dx^2+q(x,t)$,
$P_{2n+1}(t)$ a differential expression of order $2n+1$, whose
coefficients are certain differential polynomials in $q(x,t)$ (i.e.,
polynomials in $q$ and its $x$-derivatives), and $t\in \bb R$ an
additional deformation parameter. Varying $n\in \bb N\cup\{0\}$, the
collection of all Lax equations
\begin{equation} \label{1.4}
\dfrac{d}{dt}L_2 =\left[P_{2n+1},L_2 \right], \
\text{that is, }\,
q_t=\left[P_{2n+1},L_2 \right]
\end{equation}
then defines the celebrated KdV hierarchy. In particular, $q(x,t)$ is
called an algebro-geometric solution of (one of) the $n_0^{\rm th}$
equation in \eqref{1.4} if it satisfies for some (and hence for all)
fixed $t_0
\in \bb R$ one of the higher-order stationary KdV equations in
\eqref{1.4} associated with some $n_1 \geq n_0$. Therefore, without
loss
of generality, one can focus on characterizing stationary elliptic
algebro-geometric solutions of the KdV hierarchy (and similarly in
connection with other hierarchies of soliton equations).

The stationary KdV hierarchy, characterized by $q_t=0$ or
$[P_{2n+1},L_2]=0$, is intimately connected with the question of
commutativity of ordinary differential expressions. In particular, if
$[P_{2n+1},L]=0$, a celebrated theorem of Burchnall and Chaundy
\cite{11}, \cite{12} implies that $P_{2n+1}$ and $L_2$ satisfy an
algebraic relationship of the form
\begin{equation}\label{1.5}
P_{2n+1}^2 = \displaystyle \prod^{2n}_{m=0}(L_2 -E_m) \,
\text{for some} \, \{E_m\}^{2n}_{m=0} \subset \bb C
\end{equation}
and hence define a (possibly singular) hyperelliptic curve (branched at
infinity)
\begin{equation}\label{1.6}
w^2=\prod^{2n}_{m=0}(E-E_m).
\end{equation}
It is the curve (\ref{1.6}), which signifies that $q$ in $L_2
=d^2/dx^2+q(x)$ represents an algebro-geometric KdV potential.

While these considerations pertain to general solutions of the
stationary KdV hierarchy, we now concentrate on the additional
restriction that $q$ be an elliptic function (i.e., meromorphic and
doubly periodic) and hence return to the history of elliptic
algebro-geometric potentials $q$ for $L_2 = d^2/dx^2+q(x)$, or,
equivalently, elliptic solutions of the stationary KdV hierarchy.
Ince's
remarkable algebro-geometric result (\ref{1.3}) remained the only
explicit elliptic algebro-geometric example until the KdV flow
$q_t=\dfrac{1}{4}q_{xxx}+
\dfrac{3}{2}qq_x$ with the initial condition
$q(x,0)=-6\wp(x)$ was explicitly integrated by Dubrovin and Novikov
\cite{17} in 1975 (see also \cite{18}--\cite{20}, \cite{38}), and found
to be of the type
\begin{equation}\label{1.7}
q(x,t) =- 2\sum^3_{j=1} \wp(x-x_j(t))
\end{equation}
for appropriate $\{x_j(t)\}_{1\le j\le 3}$. Given these results it was
natural to ask for a systematic account of all elliptic solutions of the
KdV hierarchy, a problem posed, for instance, in \cite{49}, p. 152.

In 1977, Airault, McKean and Moser, in their seminal paper \cite{2},
presented the first systematic study of the isospectral torus $I_{\bb
R}(q_0)$ of real-valued smooth potentials $q_0(x)$ of the type
\begin{equation}\label{1.8}
q_0(x) =-2\sum^M_{j=1} \wp(x-x_j)
\end{equation}
with an algebro-geometric spectrum of the form (\ref{1.3}). In
particular, the potential (\ref{1.8}) turned out to be intimately
connected with completely integrable many-body systems of the
Calogero-Moser-type \cite{13a}, \cite{47a} (see also \cite{13c},
\cite{13b}). This connection with integrable particle systems was
subsequently exploited by Krichever \cite{44} in his fundamental
construction of elliptic algebro-geometric solutions of the
Kadomtsev-Petviashvili equation. The next breakthrough occurred in 1988
when Verdier \cite{65} published new explicit examples of elliptic
algebro-geometric potentials. Verdier's examples spurred a flurry of
activities and inspired Belokolos and Enol'skii \cite{8}, Smirnov
\cite{58}, and subsequently Taimanov \cite{59} and Kostov and Enol'skii
\cite{40} to find further such examples by combining the reduction
process of Abelian integrals to elliptic integrals (see \cite{5},
\cite{6}, \cite{6a}, Ch. 7, \cite{7}) with the aforementioned techniques
of Krichever \cite{44}, \cite{44a}. This development finally culminated
in a series of recent results of Treibich and Verdier \cite{61},
\cite{62}, \cite{63}, where it was shown that a general complex-valued
potential of the form
\begin{equation}\label{1.9}
q(x)=-\sum^4_{j=1}d_j \; \wp(x-\omega_j)
\end{equation}
$(\omega_2 =\omega_1+\omega_3, \; \omega_4=0)$ is an algebro-geometric
potential if and only if $d_j/2$ are triangular numbers, that is, if
and
only if
\begin{equation}\label{1.10}
d_j=g_j(g_j+1) \text{ for some } g_j\in \bb Z,
\quad 1\le j\le 4.
\end{equation}
We shall refer to potentials of the form \eqref{1.9}, \eqref{1.10} as
Treibich-Verdier potentials. The methods of Treibich and Verdier are
based on hyperelliptic tangent covers of the torus $\bb C/\Lambda$
($\Lambda$ being the period lattice generated by $2\omega_1$ and
$2\omega_3$).

The state of the art of elliptic algebro-geometric solutions up to 1993
was recently reviewed in issues 1 and 2 of volume 36 of Acta Applicandae
Math., see, for instance, \cite{8a}, \cite{20a}, \cite{44b}, \cite{58a},
\cite{59a}, \cite{60a} and also in \cite{8b}, \cite{13f}, \cite{14a},
\cite{20b}, \cite{32b}, \cite{55a}, \cite{58b}, \cite{60b}. In addition
to these investigations on elliptic solutions of the KdV hierarchy, the
study of other soliton hierarchies, such as the modified KdV hierarchy,
nonlinear Schr\"odinger hierarchy, and Boussinesq hierarchy has also
begun. We refer, for instance, to \cite{13d}, \cite{17a}, \cite{29},
\cite{30}, \cite{44c}, \cite{45a}, \cite{46a}, \cite{55b}, \cite{55c},
\cite{58c}, \cite{58d}, \cite{58e}, \cite{58f}.

Despite these (basically algebro-geometric) approaches described thus
far, an efficient characterization of all elliptic solutions of the KdV
hierarchy remained elusive until recently. The final breakthrough in
this characterization problem in \cite{32c}, \cite{32a} became possible
due to the application of the most powerful analytic tool in this
context, a theorem of Picard. This result of Picard (cf. Theorem
\ref{t5.1}) is concerned with the existence of solutions which are
elliptic of the second kind of $n^{\rm th}$-order ordinary differential
equations with elliptic coefficients. The main hypothesis in Picard's
theorem for a second-order differential equation of the form
\begin{equation} \label{1.11}
y''(x)+q(x)y(x)=Ey(x), \quad E\in\bb C,
\end{equation}
with an elliptic potential $q$, relevant in connection with the KdV
hierarchy (cf. the second-order differential expression $L_2$ in
\eqref{1.4}), assumes the existence of a fundamental system of solutions
meromorphic in $x$. Hence we call any elliptic function $q$ which has
this property for all values of the spectral parameter a {\bf Picard-KdV
potential}. The characterization of all elliptic algebro-geometric
solutions of the stationary KdV hierarchy, then reads as follows:
\begin{thm} \label{t1.1} (\cite{32c}, \cite{32a})
$q$ is an elliptic algebro-geometric potential if and only if it is a
Picard-KdV potential.
\end{thm}

In particular, Theorem \ref{t1.1} sheds new light on Picard's theorem
since it identifies the elliptic coefficients $q$ for which there
exists
a meromorphic fundamental system of solutions of
\eqref{1.11}
precisely as the elliptic algebro-geometric solutions of the stationary
KdV hierarchy. Moreover, we stress its straightforward applicability
based on an elementary Frobenius-type analysis which decides whether or
not \eqref{1.11} has a meromorphic fundamental system for each $E\in\bb
C$. Related results and further background information on our approach
can be found in \cite{27a}, \cite{29}--\cite{31}, \cite{32}, \cite{32d}.

After this somewhat detailed description of the history of the problem
under consideration, we now turn to the content of the present paper.
The principal objective in this paper is to prove an analogous
characterization of all elliptic algebro-geometric solutions of the
AKNS
hierarchy and hence to extend the preceding formalism to matrix-valued
differential expressions. More precisely, replace the scalar
second-order differential equation \eqref{1.2} by the $2 \times 2$
first-order system
\begin{equation} \label{1.12}
J\Psi'(E,x) + Q(x)\Psi(E,x) = E\Psi(E,x),
\quad E \in \bb C,
\end{equation}
where $\Psi(E,x)=(\psi_1(E,x),\psi_2(E,x))^t$ ($``t$'' abbreviating
transpose) and
\begin{align}
J&=\begin{pmatrix}i&0\\0&- i\end{pmatrix},
\label{1.13} \\
Q(x)&=\begin{pmatrix}Q_{1,1}(x)&Q_{1,2}(x)
\\Q_{2,1}(x)&Q_{2,2}(x)\end{pmatrix}
=\begin{pmatrix}0&-iq(x)\\ip(x)&0\end{pmatrix}.
\label{1.14}
\end{align}
Similarly, replace the scalar KdV differential expression $L_2$ by the
$2 \times 2$ matrix-valued differential expression $L(t)=J d/dx+
Q(x,t)$, $t$ a deformation parameter. The AKNS hierarchy of nonlinear
evolution equations is then constructed via appropriate Lax pairs
$(P_{n+1}(t), L(t))$, where $P_{n+1}(t)$ is a $2 \times 2$
matrix-valued
differential expression of order $n+1$ (cf. Section \ref{akns} for an
explicit construction of $P_{n+1}$). In analogy to the KdV hierarchy,
varying $n\in \bb N \cup \{0\}$, the collection of all Lax equations
\begin{equation} \label{1.15}
\dfrac{d}{dt}L =\left[P_{n+1},L \right],
\end{equation}
then defines the AKNS hierarchy of nonlinear evolution equations for
$(p(x,t), q(x,t))$. Algebro-geometric AKNS solutions are now introduced
as in the KdV context and stationary AKNS solutions, characterized by
$p_t =0, q_t=0$ or $[P_{n+1},L]=0$, again yield an algebraic
relationship
between $P_{n+1}$ and $L$ of the type
\begin{equation} \label{1.16}
P_{n+1}^2 = \prod_{m=0}^ {2n+1} (L-E_m) \,
\text{for some} \, \{E_m\}_{m=0}^{2n+1}
\subset \bb C
\end{equation}
and hence a (possibly singular) hyperelliptic curve (not branched at
infinity)
\begin{equation}\label{1.17}
w^2=\prod^{2n+1}_{m=0}(E-E_m).
\end{equation}

In order to characterize all elliptic solutions of the AKNS hierarchy
we
follow our strategy in the KdV context and consider $2 \times 2$
first-order systems of the form
\begin{equation} \label{1.18}
J\Psi'(x) + Q(x)\Psi(x) = E\Psi(x), \quad E \in\bb C,
\end{equation}
with $Q$ an elliptic $2\times 2$ matrix of the form \eqref{1.14}. Again
we single out those elliptic $Q$ such that \eqref{1.18} has a
fundamental system of solutions meromorphic in $x$ for all values of
the
spectral parameter $E\in \bb C$ and call such $Q$ {\bf Picard-AKNS
potentials}. Our principal new result in this paper, a characterization
of all elliptic algebro-geometric solutions of the stationary AKNS
hierarchy, then simply reads as follows:
\begin{thm} \label{t1.2}
An elliptic potential $Q$ is an algebro-geometric AKNS potential if and
only if it is a Picard-AKNS potential (i.e., if and only if for
infinitely many and hence for all $E\in \bb C$, \eqref{1.18} has a
fundamental system of solutions meromorphic with respect to $x$).
\end{thm}

The proof of Theorem \ref{1.2} in Section \ref{picard} (Theorem
\ref{t5.4}) relies on three main ingredients:  A purely Floquet
theoretic part to be discussed in detail in Sections \ref{floquet} and
\ref{fingap}, the fact that meromorphic algebro-geometric AKNS
potentials are Picard potentials using gauge transformations in Section
\ref{gauge}, and an elliptic function part described in Section
\ref{picard}. The corresponding Floquet theoretic part is summarized in
Theorems \ref{t3.7}, \ref{t3.8}, \ref{t4.1}--\ref{t4.4}. In particular,
Theorems \ref{t3.7} and \ref{t3.8} illustrate the great variety of
possible values of algebraic multiplicities of (anti)periodic and
Dirichlet eigenvalues in the general case where $L$ is non-self-adjoint.
Theorem \ref{t4.1} on the other hand reconstructs the (possibly
singular) hyperelliptic curve \eqref{1.17} associated with the
$2 \times
2$ periodic matrix $Q$ (not necessarily elliptic), which gives rise to
two linearly independent Floquet solutions of $J\Psi'+Q\Psi=E\Psi$ for
all but finitely many values of $E \in \bb C$.

Our use of gauge transformations in Section \ref{gauge}, in principle,
suggests a constructive method to relate $\tau$-functions associated
with a singular curve $\calK_n$ and $\theta$-functions of the associated
desingularized curve $\hat \calK_{\hat n}$, which appears to be of
independent interest.

The elliptic function portion in Section \ref{picard} consists of several
items. First of all we describe a matrix generalization of Picard's
(scalar) result in Theorem \ref{t5.1}. In Theorem \ref{t5.3} we prove
the key result that all $4\omega_j$-periodic eigenvalues associated with
$Q$ lie in certain strips
\begin{equation}\label{1.19}
S_j=\{E\in\bbC\, | \, |\Im(|\omega_j|^{-1}\omega_j E)| \leq C_j\},
\quad j=1,3
\end{equation}
for suitable constants $C_j>0$. Then $S_1$ and $S_3$ do not intersect
outside a sufficiently large disk centered at the origin. A combination
of this fact and Picard's Theorem \ref{t5.1} then yields a proof of
Theorem \ref{1.2} (see the proof of Theorem \ref{t5.4}).

We close Section \ref{picard} with a series of remarks that put
Theorem \ref{1.2} into proper perspective: Among a variety of points, we
stress, in particular, its straightforward applicability based on an
elementary Frobenius-type analysis, its property of complementing
Picard's original result, and its connection with the Weierstrass theory
of reduction of Abelian to elliptic integrals. Finally, Section \ref{Ex}
rounds off our presentation with a few explicit examples.

The result embodied by Theorems \ref{t1.1} and \ref{t1.2} in the context
of the KdV and AKNS hierarchies, uncovers a new general principle in
connection with elliptic algebro-geometric solutions of completely
integrable systems: The existence of such solutions appears to be in a
one-to-one correspondence with the existence of a meromorphic (with
respect to the independent variable) fundamental system of solutions for
the underlying linear Lax differential expression (for all values of the
corresponding spectral parameter $E\in \bb C$).

Even though the current AKNS case is technically much more involved than
the KdV case in \cite{32a} (and despite the large number of references
at
the end) we have made every effort to keep this presentation
self-contained.

\section{The AKNS Hierarchy, Recursion Relations, and
Hyperelliptic Curves} \label{akns}
\setcounter{equation}{0}
In this section we briefly review the construction of the AKNS hierarchy
using a recursive approach. This method was originally introduced by
Al'ber \cite{4} in connection with the Korteweg-de Vries hierarchy. The
present case of the AKNS hierarchy was first systematically developed in
\cite{27b}.

Suppose that $q=iQ_{1,2}, \, p=-iQ_{2,1} \in C^{\infty} (\bb R)$ (or
meromorphic on $\bb C$) and consider the Dirac-type matrix-valued
differential expression
\begin{equation} \label{2.1}
L=J\frac{d}{dx} + Q(x)
=\begin{pmatrix}i&0\\0&-i\end{pmatrix}\frac{d}{dx}+
\begin{pmatrix}0&-iq(x)\\ip(x)&0\end{pmatrix},
\end{equation}
where we abbreviate
\begin{align}
J&=\begin{pmatrix}i&0\\0&- i\end{pmatrix},\label{2.2} \\
Q(x)&=\begin{pmatrix}Q_{1,1}(x)&Q_{1,2}(x)
\\Q_{2,1}(x)&Q_{2,2}(x)\end{pmatrix}
=\begin{pmatrix}0&-iq(x)\\ip(x)&0\end{pmatrix}.
\label{2.3}
\end{align}
In order to explicitly construct higher-order matrix-valued differential
expressions $P_{n+1}$, $n\in\bb N_0$ (=$\bb N \cup \{0\}$) commuting
with $L$, which will be used to define the stationary AKNS hierarchy,
one can proceed as follows (see \cite{27b} for more details).

Define functions $f_\ell$, $g_\ell$, and $h_\ell$ by the following
recurrence relations,
\begin{gather}
  f_{-1}=0,\quad g_0=1,\quad h_{-1}=0,\notag \\
  f_{\ell+1}=\frac{i}{2} f_{\ell,x} - iqg_{\ell+1}, \quad
  g_{\ell+1,x}=pf_{\ell}+q h_{\ell}, \quad
  h_{\ell+1}=-\frac{i}{2}h_{\ell,x} + ip g_{\ell+1}
\label{2.4}
\end{gather}
for $\ell=-1,0,1,...$. The functions $f_\ell$, $g_\ell$, and $h_\ell$
are polynomials in the variables $p,q,p_x,q_x,...$ and $c_1,c_2,...$
where the $c_j$ denote integration constants. Assigning weight $k+1$ to
$p^{(k)}$ and $q^{(k)}$ and weight $k$ to $c_k$ one finds that $f_\ell$,
$g_{\ell+1}$, and $h_\ell$ are homogeneous of weight $\ell+1$.

Explicitly, one computes,
\begin{align}
 &f_0 = -iq,\notag \\
 &f_1 =\frac{1}{2} q_x+c_1(-iq),\notag \displaybreak[0]\\
 &f_2=\frac{i}{4} q_{xx}-\frac{i}{2}pq^2
 +c_1( \frac{1}{2} q_x)+c_2(-iq),\notag \displaybreak[2]\\
&g_0 = 1,\notag \\
&g_1 =c_1,\notag \\
&g_2 =\frac{1}{2} pq+c_2,\notag \\
&g_3 =  -\frac{i}{4}(p_{_x}q - pq_x) +
c_1(\frac{1}{2} pq)
+ c_3 , \label{2.5}\\
&h_0 =i p,\notag \\
&h_1 =\frac{1}{2} p_{_x}+ c_1(ip), \notag \\
&h_2 =-\frac{i}{4} p_{_{xx}}+\frac{i}{2}p{^2}q+
c_1( \frac{1}{2}
p_{_x})+c_2(i p), \notag \\
& \text{etc}. \notag
\end{align}

Next one defines the matrix-valued differential expression $P_{n+1}$ by
\begin{equation} \label{2.6}
P_{n+1}=-\sum_{\ell=0}^{n+1}(g_{n-\ell+1}J+iA_{n-\ell})L^{\ell},
\end{equation}
where
\begin{equation} \label{2.7}
A_{\ell}=\begin{pmatrix}0&-f_{\ell}
\\h_{\ell}&0\end{pmatrix},\quad \ell=-1,0,1,\dots .
\end{equation}
One verifies that
\begin{equation} \label{2.8}
[g_{n-\ell+1}J+iA_{n-\ell},L]=2iA_{n-\ell}L-2iA_{n-\ell+1},
\end{equation}
where $[\cdot,\cdot]$ denotes the commutator.
This implies
\begin{equation} \label{2.9}
[P_{n+1},L]=2i A_{n+1}.
\end{equation}
The pair $(P_{n+1},L)$ represents a Lax pair for the AKNS hierarchy.
Introducing a deformation parameter $t$ into $(p,q)$, that is,
$(p(x),q(x))
\rightarrow(p(x,t),q(x,t))$, the AKNS hierarchy (cf., e.g., \cite{47b},
Chs. 3, 5 and the references therein) is defined as the collection of
evolution equations (varying $n \in \bbN_0$)
\begin{equation}\label{2.40}
\frac{d}{dt}L(t)-[P_{n+1}(t),L(t)]=0
\end{equation}
or equivalently, by
\begin{equation}\label{2.41}
\mbox{AKNS}_n(p,q)=\begin{pmatrix}p_{t}(x,t)-2 h_{n+1}(x,t)\\
q_{t}(x,t)-2 f_{n+1}(x,t)\end{pmatrix}=0,
\end{equation}
that is, by
\begin{equation}\label{2.42}
\mbox{AKNS}_n(p,q)=\begin{pmatrix}p_{t}+i(H_{n,x}-2iE H_n-2p G_{n+1})\\
q_{t}-i(F_{n,x}+2iE F_n-2q G_{n+1})\end{pmatrix}=0.
\end{equation}

Explicitly, one obtains for the first few equations in (\ref{2.41})
\begin{align}
&\mbox{AKNS}_0(p,q)=\begin{pmatrix} p_{t}-p_{x}-2i c_1p\\
    q_{t} - q_x + 2i c_1q\end{pmatrix}=0 \notag \\
&\mbox{AKNS}_1(p,q)=\begin{pmatrix}
 p_{t}+\frac{i}{2} p_{xx}-i p^2q-c_1 p_x-2i c_2 p  \\
 q_{t}-\frac{i}{2} q_{xx}+i pq^2-c_1 q_x+2i c_2 q \end{pmatrix}=0
 \label{2.43}\\
&\mbox{AKNS}_2(p,q)=\begin{pmatrix}
 p_{t}+\frac14p_{xxx}-\frac32pp_xq+c_1(\frac{i}2p_{xx}- i p^2q)
  -c_2 p_x-2i c_3 p \\
 q_{t}+\frac14q_{xxx}-\frac32pqq_x+c_1(-\frac{i}2q_{xx}+i pq^2)
  -c_2 q_x+2i c_3 q \end{pmatrix} =0 \notag \\
&\text{etc}. \notag
\end{align}

The stationary AKNS hierarchy is then defined by the vanishing of the
commutator of $P_{n+1}$ and $L$, that is, by
\begin{equation} \label{2.10}
[P_{n+1},L]=0, \quad n\in \bb N_0,
\end{equation}
or equivalently, by
\begin{equation} \label{2.11}
f_{n+1}=h_{n+1}=0, \quad n\in \bb N_0.
\end{equation}

Next, we introduce $F_n$, $G_{n+1}$, and $H_n$ which are polynomials
with respect to $E\in \bb C$,
\begin{align}
F_n(E,x)&= \sum_{\ell=0}^{n}f_{n-\ell}(x)E^{\ell},\notag \\
G_{n+1}(E,x)&= \sum_{\ell=0}^{n+1} g_{n+1-\ell}(x)E^\ell,\label{2.13}\\
H_n(E,x)&= \sum_{\ell=0}^{n}h_{n-\ell}(x)E^\ell,\notag
\end{align}
and note that \eqref{2.11} becomes
\begin{align}
F_ {n,x}(E,x)&=-2 i E F_n(E,x)+2q(x) G_{n+1}(E,x),\label{2.14}\\
G_{n+1,x}(E,x)&=p(x)F_n(E,x)+q(x)H_n(E,x), \label{2.15}\\
H_{n,x}(E,x)&=2iE H_n(E,x)+2 p(x) G_{n+1}(E,x).
\label{2.16}
\end{align}
These equations show that $G_{n+1}^2 - F_nH_n$ is independent of $x$.
Hence
\begin{equation} \label{2.17}
R_{2n+2}(E)=G_{n+1}(E,x)^2- F_n(E,x)H_n(E,x)
\end{equation}
is a monic polynomial in $E$ of degree $2n+2$.

One can use \eqref{2.14}--\eqref{2.17} to derive differential equations
for $F_n$ and $H_n$ separately by eliminating $G_{n+1}$. One obtains
\begin{align}
&q(2F_nF_{n,xx}-F_{n,x}^2+4(E^2-pq)F_n^2)
-q_x(2F_nF_{n,x}+4iE F_n^2)=-4 q^3 R_{2n+2}(E),\label{2.18} \\
&p(2H_nH_{n,xx}-H_{n,x}^2+4(E^2-pq)H_n^2)
-p_x(2H_nH_{n,x}-4iE H_n^2)=-4 p^3 R_{2n+2}(E).\label{2.19}
\end{align}

Next, assuming $[P_{n+1},L]=0$, one infers
\begin{equation} \label{2.20}
P_{n+1}^2=\sum_{\ell,m=0}^{n+1}(g_{n-\ell+1}J+iA_{n-\ell})
(g_{n-m+1}J+iA_{n-m})L^{\ell+m}.
\end{equation}
Hence,
\begin{equation} \label{2.21}
P_{n+1}^2=-G_{n+1}(L,x)^2+F_n(L,x)H_n(L,x)=-R_{2n+2}(L),
\end{equation}
that is, whenever $P_{n+1}$ and $L$ commute they necessarily
satisfy an
algebraic relationship. In particular, they define a (possibly
singular)
hyperelliptic curve $\calK_n$ of (arithmetic) genus $n$ of the type
\begin{equation} \label{2.22}
\calK_n: \, w^2 = R_{2n+2}(E), \quad R_{2n+2}(E)=
\prod_{m=0}^ {2n+1} (E-E_m) \, \text{for some} \,
\{E_m\}_{m=0}^{2n+1}
\subset \bb C.
\end{equation}
The functions $f_\ell$, $g_\ell$, and $h_\ell$, and hence the matrices
$A_\ell$ and the differential expressions $P_{\ell}$ defined above,
depend on the choice of the integration constants $c_1,c_2,...,c_\ell$
(cf. \eqref{2.5}). In the following we make this dependence
explicit and write $f_{\ell}(c_1,\dots,c_{\ell})$,
$g_{\ell}(c_1,\dots,c_{\ell})$, $h_{\ell}(c_1,\dots,c_{\ell})$,
$A_\ell(c_1,..., c_\ell)$, $P_\ell(c_1,...,c_\ell)$, etc. In particular,
we denote homogeneous quantities, where $c_{\ell}=0, \ell
\in \bb N$ by $\hat f_{\ell} = f_{\ell}(0,\dots,0)$,
$\hat g_{\ell} = g_{\ell}(0,\dots,0)$,
$\hat h_{\ell} = h_{\ell}(0,\dots,0)$,
$\hat A_{\ell} = A_{\ell}(0,\dots,0)$,
$\hat P_\ell=P_\ell(0,...,0)$, etc. In addition, we
note that
\begin{equation} \label{2.23}
f_\ell(c_1,...,c_\ell) =\sum_{k=0}^\ell
c_{\ell-k}\hat f_k, \quad
g_\ell(c_1,...,c_\ell) =\sum_{k=0}^\ell
c_{\ell-k}\hat g_k, \quad
h_\ell(c_1,...,c_\ell) =\sum_{k=0}^\ell
c_{\ell-k}\hat h_k,
\end{equation}
and
\begin{equation} \label{2.24}
A_\ell(c_1,...,c_\ell) =\sum_{k=0}^\ell c_{\ell-k}\hat A_k,
\end{equation}
defining $c_0=1$. In particular, then
\begin{equation} \label{2.26}
P_r(c_1,...,c_r)=\sum_{\ell=0}^r c_{r-\ell} \hat P_{\ell}.
\end{equation}

Next suppose that $P_{n+1}$ is any $2 \times 2$ matrix-valued
differential expression such that $[P_{n+1},L]$ represents
multiplication by a matrix whose diagonal entries are zero. This implies
that the leading coefficient of $P_{n+1}$ is a constant diagonal matrix.
Since any constant diagonal matrix can be written as a linear
combination of $J$ and $I$ (the identity matrix in $\bb C^2$), we infer
the existence of complex numbers $\alpha_{n+1}$ and $\beta_{n+1}$ such
that
\begin{equation} \label{2.27}
S_1=P_{n+1} -\alpha_{n+1}\hat P_{n+1}-\beta_{n+1}L^{n+1}
\end{equation}
is a differential expression of order at most $n$ whenever $P_{n+1}$ is
of order $n+1$. Note that
\begin{equation} \label{2.28}
[S_1,L]=[P_{n+1},L]-\alpha_{n+1}[\hat P_{n+1},L]=
[P_{n+1},L]-2i\alpha_{n+1}\hat A_{n+1}
\end{equation}
represents multiplication with zero diagonal elements. An induction
argument then shows that there exists $S_{n+1}$ such that
\begin{equation} \label{2.30}
S_{n+1}=P_{n+1} -\sum_{\ell=1}^{n+1}(\alpha_{\ell}
\hat P_{\ell}+\beta_{\ell} L^{\ell})
\text{ and }
[S_{n+1},L]=[P_{n+1},L]-2i\sum_{\ell=1}^{n+1}
\alpha_{\ell} \hat A_{\ell}.
\end{equation}
Since the right-hand side of the last equation is multiplication with a
zero diagonal, $S_{n+1}$ is a constant diagonal matrix, that is, there
exist complex numbers $\alpha_0$ and $\beta_0$ such that
$S_{n+1}=\alpha_0J+\beta_0I$. Hence
\begin{equation} \label{2.31}
P_{n+1}=\sum_{\ell=0}^{n+1}(\alpha_{\ell} \hat P_{\ell}
+\beta_{\ell} L^{\ell})
\text{ and }
[P_{n+1},L]=2i\sum_{\ell=0}^{n+1}\alpha_{\ell}\hat A_{\ell}.
\end{equation}
Consequently, if all $\alpha_{\ell}=0$, then $P_{n+1}$ is a polynomial
of $L$, and $P_{n+1}$ and $L$ commute irrespective of $p$ and $q$. If,
however, $\alpha_r\neq0$ and $\alpha_{\ell}=0$ for $\ell >r$, then
\begin{equation} \label{2.32}
P_{n+1}=\alpha_r P_r(\frac{\alpha_{r-1}}{\alpha_r},...,
\frac{\alpha_0}{\alpha_r})+\sum_{\ell=0}^{n+1}\beta_{\ell} L^{\ell}.
\end{equation}
In this case $P_{n+1}$ and $L$ commute if and only if
\begin{equation} \label{2.33}
\sum_{\ell=0}^r \frac{\alpha_{\ell}}{\alpha_r}\hat A_{\ell}
=A_r(\frac{\alpha_{r-1}}{\alpha_r},...,\frac{\alpha_0}{\alpha_r})=0,
\end{equation}
that is, if and only if $(p,q)$ is a solution of some equation of the
stationary AKNS hierarchy. In this case $P_{n+1}-\sum_{\ell=0}^{n+1}
\beta_{\ell} L^{\ell}$ and $L$ satisfy an algebraic relationship of the
type \eqref{2.21}.

Suppose on the other hand that $P_{n+1}$ is a matrix-valued differential
expression such that $(P_{n+1}-K_r(L))^2=-R_{2n+2}(L)$ for some
polynomials $K_r$ and $R_{2n+2}$. Then $L$ commutes with
$(P_{n+1}-K_r(L))^2$ and this enforces that $L$ also commutes with
$P_{n+1}-K_r(L)$ and hence with $P_{n+1}$. Thus we proved the following
theorem.
\begin{thm} \label{t2.1}
Let $L$ be defined as in \eqref{2.1}. If $P_{n+1}$ is a matrix-valued
differential expression of order $n+1$ which commutes with $L$, whose
leading coefficient is different from a constant multiple of $J^{n+1}$,
then there exist polynomials $K_r$ and $R_{2n+2}$ of degree $r\leq n+1$
and $2n+2$, respectively, such that $(P_{n+1}-K_r(L))^2=-R_{2n+2}(L)$.
\end{thm}

Theorem \ref{t2.1} represents a matrix-valued generalization of a
celebrated result due to Burchnall and Chaundy \cite{11}, \cite{12} in
the special case of scalar differential expressions.

By the arguments presented thus far in this section it becomes natural
to make the following definition. We denote by $M_2 (\bb C)$ the set of
all $2\times 2$ matrices over $\bb C$.
\begin{defn} \label{d2.2}
A function $Q:\bb R\to M_2 (\bb C)$ of the type
$Q=\begin{pmatrix}0&-iq\\ip&0\end{pmatrix}$ is called an {\bf
algebro-geometric AKNS potential} if $(p,q)$ is a stationary solution of
some equation of the AKNS hierarchy \eqref{2.11}.
\end{defn}

By a slight abuse of notation we will also call $(p,q)$ an
algebro-geometric AKNS potential in this case.

The following theorem gives a sufficient condition for $Q$ to be
algebro-geometric.
\begin{thm} \label{t2.3}
Assume that $F_n(E,x)=\sum_{\ell=0}^n f_{n-\ell}(x)E^\ell$ with
$f_0(x)=-iq(x)$ is a polynomial of degree $n$ in $E$, whose coefficients
are twice continuously differentiable complex-valued functions on
$(a,b)$ for some $-\infty \leq a<b \leq \infty$. Moreover, suppose that
$q$ has (at most) finitely many zeros on each compact interval on $\bb
R$ with $\pm \infty$ the only possible accumulation points. If
\begin{align}
\frac1{4q(x)^3}\Bigl\{&q(x)\bigl(2F_n(E,x)F_{n,xx}(E,x)
-F_{n,x} (E,x)^2+4(E^2-p(x)q(x)) F_n(E,x)^2\bigr) \notag \\
&-q_x (x)\bigl(2F_n(E,x)F_{n,x} (E,x)+
4iE F_n(E,x)^2\bigr)\Bigr\}, \label{2.33a}
\end{align}
is independent of $x$, then $p$, $q$ and all coefficients $f_{\ell}$ of
$F_n$ are in $C^{\infty}((a,b))$. Next, define
\begin{align}
g_0(x)&=1, \notag \\ g_{\ell+1}(x)&=\frac{1}{q(x)} (-\frac12
f_{\ell,x}(x)-if_{\ell+1}(x)),\label{2.37} \\
h_\ell(x)&=\frac{1}{q(x)}(-g_{\ell+1,x}(x)- ip(x)f_\ell(x))
\label{2.38}
\end{align}
for $\ell=0, \dots ,n$, where $f_{n+1}=0$. Then the differential
expression $P_{n+1}$ defined by \eqref{2.6} commutes with $L$ in
\eqref{2.1}. In particular, if $(a,b)=\bb R$ then $(p,q)$ is an
algebro-geometric AKNS potential.
\end{thm}

\begin{proof}
The expression \eqref{2.33a} is a monic polynomial of degree $2n+2$ with
constant coefficients. We denote it by
\begin{equation} \label{2.34}
R_{2n+2}(E)=\sum_{m=0}^{2n+2}\gamma_{2n+2-m}E^m,\quad \gamma_0 =1.
\end{equation}
We now compare coefficients in \eqref{2.33a} and \eqref{2.34} starting
with the largest powers. First of all this yields
\begin{equation} \label{2.35}
q_x=i\gamma_1 q-2 f_1,
\end{equation}
which shows that $q\in C^3((a,b))$, and secondly that
\begin{equation} \label{2.36}
p=(-4\gamma_2 q^2-3q_x^2+2qq_{xx}-8iqf_2-4f_1^2+8q_xf_1)/(4q^3),
\end{equation}
which shows that $p\in C^1((a,b))$. Comparing the coefficients of
$E^{2n-\ell}$ allows one to express $4q^2f_{\ell,xx}$ as a polynomial in
$p$, $q$, $q_x$, the coefficients $f_{\ell}$, their first derivatives,
and the second derivatives of $f_0$,...,$f_{\ell-1}$. Therefore, one may
show recursively that $f_{\ell,xx}\in C^1((a,b))$ for any
$\ell\in\{1,...,n\}$. Equations \eqref{2.35} and \eqref{2.36} then show
that $q\in C^4((a,b))$ and $p\in C^2((a,b))$. Thus it follows that the
$f_{\ell,xx}$ are in $C^2((a,b))$. An induction argument now completes
the proof of the first part of the theorem.

Next, introducing
\begin{equation} \label{2.39}
G_{n+1}(E,x)=\sum_{\ell=0}^{n+1}g_{n+1-\ell}(x) E^\ell,\quad
H_n(E,x)=\sum_{\ell=0}^{n}h_{n-\ell}(x) E^\ell,
\end{equation}
one finds that $F_n$, $G_{n+1}$, and $H_n$ satisfy equations
\eqref{2.14} and \eqref{2.15}. Equating \eqref{2.33a} with $R_{2n+2}$
yields $R_{2n+2}=G_{n+1}^2-F_nH_n$. Hence $G_{n+1}^2-F_nH_n$ does not
depend on $x$ and therefore differentiating with respect to $x$ results
in $2G_{n+1}G_{n+1,x}-F_nH_{n,x}-F_{n,x}H_n=0$ which shows that
equation \eqref{2.16} also holds. This in turn proves that $f_\ell$,
$g_\ell$, and $h_\ell$ satisfy the recurrence relations given above with
$f_{n+1}=h_{n+1}=0$. The commutativity of $P_{n+1}$ and $L$ now follows
as before.
\end{proof}

The same proof yields the following result.

\begin{cor} \lb{c2.3a}
Assume that $F_n(E,x)=\sum_{\ell=0}^n f_{n-\ell}(x)E^\ell$,
$f_0(x)=-iq(x)$ is a polynomial of degree $n$ in $E$ whose coefficients
are meromorphic in $x$. If
\begin{align}
\frac1{4q(x)^3}\Bigl\{&q(x)\bigl(2F_n(E,x)F_{n,xx}(E,x)
-F_{n,x} (E,x)^2+4(E^2
-p(x)q(x)) F_n(E,x)^2\bigr) \notag \\
&-q_x (x)\bigl(2F_n(E,x)F_{n,x} (E,x)+ 4iE F_n(E,x)^2\bigr)\Bigr\},
\label{2.33b}
\end{align}
is independent of $x$, then $(p,q)$ is a meromorphic algebro-geometric
AKNS potential.
\end{cor}

Finally, we mention an interesting scale invariance of the AKNS
equations \eqref{2.41}.
\begin{lem}\lb{l2.4}
Suppose $(p,q)$ satisfies one of the AKNS equations \eqref{2.41},
\ba \label{2.44}
\text{AKNS}_n(p,q)=0
\ea
for some $n\in\bbN_0$. Consider the scale transformation
\ba \label{2.45}
(p(x,t),q(x,t))\to ({\breve p}(x,t), {\breve q}(x,t))
=(A p(x,t),A^{-1}q(x,t)),\; \; A\in \bbC\backslash\{0\}.
\ea
Then
\ba \label{2.46}
\text{AKNS}_n({\breve p},{\breve q})=0.
\ea
\end{lem}
We omit the straightforward proof which can be found, for instance, in
\cite{27b}.

In the particular case of the nonlinear Schr\"{o}dinger (NS) hierarchy,
where
\ba \label{2.47}
p(x,t)=\pm \overline{q(x,t)},
\ea
\eqref{2.45} further restricts $A$ to be unimodular, that is,
\ba \label{2.48}
|A|=1.
\ea

Note that the KdV hierarchy as well as the modified Korteweg-de Vries
(mKdV) hierarchy are contained in the AKNS hierarchy. In fact, setting
all integration constants $c_{2\ell+1}$ equal to zero the $n^{\rm th}$
KdV equation is obtained from the $(2n)^{\rm th}$ AKNS system by the
constraint
\begin{equation} \label{2.51}
p(x,t) = 1
\end{equation}
while the $n^{\rm th}$ mKdV equation is obtained from the $(2n)^{\rm
th}$ AKNS system by the constraint
\begin{equation}\label{2.49}
p(x,t)=\pm q(x,t).
\end{equation}

\section{Gauge Transformations for the Stationary AKNS Hierarchy}
\label{gauge}
\setcounter{equation}{0}
This section is devoted to a study of meromorphic properties of
solutions $\Psi_{\pm}(E,x)$ of $L\Psi= E\Psi$ with respect to
$x\in\bbC$
under the assumption that $Q$ is a meromorphic algebro-geometric AKNS
potential associated with a (possibly singular) hyperelliptic curve
$\calK_n$. Meromorphic properties of $\Psi_{\pm}(E,x)$ will enter at a
crucial stage in the proof of our main characterization result,
Theorem
\ref{t5.4}.

In the following we denote the order of a meromorphic function $f$
at the
point $x\in\bbC$ by $\ord_{x} (f)$.

\begin{prop} \label{p551}
If $Q$ is a meromorphic algebro-geometric AKNS potential then
$$\ord_{x} (p)+\ord_{x} (q)\geq -2$$
for every $x\in\bb C$.
\end{prop}
\begin{proof}
Assume the contrary (i.e., $\ord_{x} (p)+\ord_{x} (q) \leq -1$) and
choose $E$ such that $R_{2n+2}(E)=0$,
where $R_{2n+2}$ is the polynomial defining the hyperelliptic curve
associated with $Q$. Define $F_n$ as in \eqref{2.13} and denote its
order at $x$ by $r$. Then $\ord_x (-4pq F_n(E,\cdot)^2)$ is strictly
smaller than $2r-2$ while the order of any other term on the left-hand
side of \eqref{2.18} is at least $2r-2$. This is impossible since the
right and thus the left-hand side of \eqref{2.18} vanishes identically.
\end{proof}

Therefore, to discuss algebro-geometric AKNS potentials we only need to
consider the case where $p$ and $q$ have Laurent expansions of the form
\begin{equation} \label{11042}
p(x)=\sum_{j=0}^\infty p_j (x-x_0)^{j-1+m},
\quad q(x)=\sum_{j=0}^\infty q_j (x-x_0)^{j-1-m},
\end{equation}
with $m$ an integer and at least one of the numbers $p_0$ and $q_0$
different from zero. If $B=\diag(-m,0)$, the change of variables
$y=x^Bw$ transforms the differential equation $Jy'+Qy=Ey$ into the
equation $w'=(R/x +S+\sum_{j=0}^\infty A_{j+1}x^j)w$, where
\begin{equation}
R=\begin{pmatrix} m&q_0\\p_0&0\end{pmatrix}, \quad
S=-iE\begin{pmatrix}1&0\\0&-1\end{pmatrix}, \quad
A_j=\begin{pmatrix}0&q_j\\p_j&0\end{pmatrix}.
\end{equation}
We now make the ansatz
\begin{equation} \label{05052}
w(x)=\sum_{j=0}^\infty \Omega_j (x-x_0)^{j+T},
\end{equation}
where $T$ and the $\Omega_j$ are suitable constant matrices. This ansatz
yields the recurrence relation
\begin{gather}
R\Omega_0-\Omega_0 T=0, \label{11040}\\
 R\Omega_{j+1}-\Omega_{j+1}(T+j+1)=-S\Omega_j
-\sum_{l=0}^j A_{l+1}\Omega_{j-l}=B_j, \label{11041}
\end{gather}
where the last equality defines $B_j$. In the following we denote the
$l^{\rm th}$ column of $\Omega_j$ and $B_j$ by $\omega_j^{(l)}$ and
$b_j^{(l)}$, respectively.

\begin{prop} \label{p552}
Suppose $Q$ is a meromorphic potential of $L\Psi=E\Psi$ and $x_0$ is a
pole of $Q$ where $p$ and $q$ have Laurent expansions given by
\eqref{11042}. The equation $L\Psi=E\Psi$ has a fundamental system of
solutions which are meromorphic in a vicinity of $x_0$ if and
only if \\
 (i) the eigenvalues $\lambda$ and $m-\lambda$ (where, without loss of
generality, $\lambda>m-\lambda$) of $R$ are distinct integers and\\
 (ii) $b_{2\lambda-m-1}$ is in the range of $R-\lambda$.
\end{prop}

\begin{proof}
A fundamental matrix of $w'=(R/x+S+\sum_{j=0}^\infty A_{j+1}x^j)w$ may
be written as \eqref{05052} where $T$ is in Jordan normal form. If  all
solutions of $L\Psi=E\Psi$ and hence of $w'=(R/x +S+\sum_{j=0}^\infty
A_{j+1}x^j)w$ are meromorphic near $x_0$, then $T$ must be a diagonal
matrix with
integer eigenvalues. Equation \eqref{11040} then shows that the
eigenvalues
of $T$ are the eigenvalues of $R$ and that $R$ is diagonalizable. But
since at least one of $p_0$ and $q_0$ is different from zero, $R$ is not
diagonalizable if $\lambda$ is a double eigenvalue of $R$, a case which
is therefore precluded. This proves (i). Since $T$ is a diagonal matrix,
equation \eqref{11041} implies
\begin{equation}
(R+\lambda-m-j-1)\omega_{j+1}^{(2)}=b_j^{(2)}
\end{equation}
for $j=0,1,...$. Statement (ii) is just the special case where
$j=2\lambda-m-1$.

Conversely, assume that (i) and (ii) are satisfied. If the recurrence
relations \eqref{11040}, \eqref{11041} are satisfied, that is, if $w$ is
a formal solution of $w'=(R/x+S+\sum_{j=0}^\infty A_{j+1}x^j)w$ then it
is also an actual solution near $x_0$ (see, e.g., Coddington and
Levinson \cite{CL}, Sect. 4.3). Since $R$ has distinct eigenvalues it
has linearly independent eigenvectors. Using these as the columns of
$\Omega_0$ and defining $T=\Omega_0^{-1}R\Omega_0$ yields
\eqref{11040}. Since $T$ is a diagonal matrix, \eqref{11041} is
equivalent to the system
\begin{align}
b_j^{(1)}&=(R-\lambda-j-1)\omega_{j+1}^{(1)}, \label{05053}\\
b_j^{(2)}&=(R+\lambda-m-j-1)\omega_{j+1}^{(2)}. \label{05054}
\end{align}
Next, we note that $R-\lambda-j-1$ is invertible for all $j\in\bb N_0$.
However,
$R+\lambda-m-j-1$ is only invertible if $j\neq 2\lambda-m-1$. Hence a
solution of the proposed form exists if and only if
$b_{2\lambda-m-1}^{(2)}$ is in the range of $R-\lambda$, which is
guaranteed
by hypothesis (ii).
\end{proof}

Note that
\begin{equation}
B_0=-E J\Omega_0-A_1\Omega_0
\end{equation}
is a first-order polynomial in $E$. As long as $R$ has distinct
eigenvalues and $j\leq 2\lambda -m-1$, we may compute $\Omega_j$
recursively from \eqref{05053} and \eqref{05054} and $B_j$ from the
equality on the right in \eqref{11041}. By induction one can show that
$\Omega_j$ is a polynomial of degree $j$ and that $B_j$ is a polynomial
of degree $j+1$ in $E$. This leads to the following result.
\begin{thm} \label{t05051}
Suppose $Q$ is a meromorphic potential of $L\Psi=E\Psi$. The equation
$L\Psi=E\Psi$ has a fundamental system of solutions which are
meromorphic with respect to the independent variable for all values of
the spectral parameter $E\in\bb C$ whenever this is true for a
sufficiently large finite number of
distinct values of $E$.
\end{thm}
\begin{proof}
By hypothesis, $Q$ has countably many poles. Let $x_0$ be any one of
them. Near $x_0$
the functions $p$ and $q$ have the Laurent expansions \eqref{11042}. The
associated matrix $R$ has eigenvalues $\lambda$ and $m-\lambda$, which
are independent of $E$. The vector $v=(q_0,-\lambda)^t$ spans
$R-\lambda$, and the determinant of the matrix whose columns are $v$ and
$b_{2\lambda-m-1}$ is a polynomial in $E$ of degree $2\lambda-m$. Our
hypotheses and Proposition \ref{p552} imply that this determinant has
more than $2\lambda-m$ zeros and hence is identically equal to zero.
This shows that $b_{2\lambda-m-1}$ is a multiple of $v$ for every value
of $E$. Applying Proposition \ref{p552} once more then shows that all
solutions of $L\Psi=E\Psi$ are meromorphic near $x_0$ for all
$E\in \bb C$. Since
$x_0$ was arbitrary, this concludes the proof.
\end{proof}

Next, let $\{E_0,...,E_{2n+1}\}$ be a set of not necessarily distinct
complex numbers. We recall (cf. (\ref{2.22})),
\begin{equation} \label{6.1}
{\calK}_n=\{P=(E,V)\, | \, V^2 = R_{2n+2}(E)=\prod_{m=0}^{2n+1}
(E-E_m)\}.
\end{equation}
We introduce the meromorphic function $\phi(\cdot,x)$ on $\calK_n$ by
\begin{equation} \label{6.2}
\phi(P,x) = \frac{V+G_{n+1}(E,x)}{F_n(E,x)},\quad P=(E,V) \in \calK_n.
\end{equation}
We remark that $\phi$ can be extended to a meromorphic function on the
compactification (projective closure) of the affine curve $\calK_n$.
This compactification is obtained by joining two points to $\calK_n$.

Next we define
\begin{align}
\psi_1(P,x,x_0) &= \exp \left\{\int_{x_0}^x\, dx'
[-iE+q(x') \phi(P,x')] \right\}, \label{6.3} \\
\psi_2 (P,x,x_0) &= \phi(P,x) \psi_1 (P,x,x_0), \label{6.4}
\end{align}
where the simple Jordan arc from $x_0$ to $x$ in (\ref{6.3}) avoids
poles of $q$ and $\phi$. One verifies with the help of
\eqref{2.14}--\eqref{2.17}, that
\begin{equation} \label{6.7}
\phi_x(P,x)=p(x)-q(x)\phi(P,x)^2+2iE\phi(P,x).
\end{equation}
 From this and \eqref{2.6} we find
\begin{equation} \label{6.5}
L \Psi(P,x,x_0) = E \Psi(P,x,x_0), \quad
P_{n+1}\Psi(P,x,x_0)=iV\Psi(P,x,x_0),
\end{equation}
where
\begin{equation} \label{6.6}
\Psi(P,x,x_0)=\begin{pmatrix}
{\psi}_1(P,x,x_0) \\{\psi}_2(P,x,x_0) \end{pmatrix}.
\end{equation}

One observes that the two branches $\Psi_{\pm} (E,x,x_0)
=(\psi_{\pm,1}(E,x,x_0),\psi_{\pm,2}(E,x,x_0))^t$ of $\Psi(P,x,x_0)$
represent a fundamental system of solutions of $Ly=Ey$ for all $E\in\bb
C\setminus \{ \{E_m\}_{m=0}^{2n+1} \cup\{\mu_j(x_0)\}_{j=1}^n \}$,
since
\begin{equation} \label{6.8}
W(\Psi_-(E,x,x_0), \Psi_+(E,x,x_0))=\frac{2V_+(E)}{F_n(E,x_0)}.
\end{equation}
Here $W(f,g)$ denotes the determinant of the two columns $f$ and $g$
and
$V_+(\cdot)$ (resp. $V_-(\cdot))$ denotes the branch of $V(\cdot)$ on
the upper (resp. lower) sheet of $\calK_n$ (we follow the notation
established in \cite{27b}).

In the special case where $\calK_n$ is nonsingular, that is, $E_m\neq
E_{m'}$ for $m\neq m'$, the explicit representation of $\Psi(P,x,x_0)$
in terms of the Riemann theta function associated with $\calK_n$
immediately proves that $\Psi_{\pm}(E,x,x_0)$ are meromorphic with
respect to $x\in \bbC$ for all $E\in \bbC \setminus
\{\{E_m\}_{m=0}^{2n+1} \cup\{\mu_j(x_0)\}_{j=1}^n\}$. A detailed account
of this theta function representation can be found, for instance, in
Theorem~3.5 of \cite{27b}. In the following we demonstrate how to use
gauge transformations to reduce the case of singular curves $\calK_n$ to
nonsingular ones.

Let $(p,q)$ be meromorphic on $\bb C$, the precise conditions on $(p,q)$
being immaterial (at least, temporarily) for introducing gauge
transformations below. Define $L$ and $Q$ as in (\ref{2.1}), (\ref{2.3})
and consider the formal first-order differential system $L\Psi=E\Psi$.
Introducing,
\begin{equation} \label{6.11}
A(E,x)=\begin{pmatrix}iE&-q(x)\\-p(x)&-iE\end{pmatrix},
\end{equation}
$L\Psi=E\Psi$ is equivalent to $\Psi_x (E,x)+A(E,x)\Psi(E,x)=0$.

Next we consider the gauge transformation,
\begin{gather}
\tilde \Psi(E,x)=\Gamma(E,x)\Psi(E,x), \label{6.13} \\
\tilde A(E,x)=\begin{pmatrix}iE& -\tilde q(x)\\
-\tilde p(x)&-iE\end{pmatrix}
=\Gamma(E,x)A(E,x)\Gamma(E,x)^{-1} -
\Gamma_x(E,x)\Gamma(E,x)^{-1}, \label{6.14}
\end{gather}
implying
\begin{equation} \label{6.15}
\tilde \Psi_x(E,x)+\tilde A(E,x)\tilde \Psi(E,x)=0, \, \,
\text{that is,} \, \, \tilde L \tilde \Psi(E,x)=E\tilde \Psi(E,x),
\end{equation}
with $\tilde L$ defined as in (\ref{3.0}), (\ref{3.1}) replacing $(p,q)$
by $(\tilde p, \tilde q)$. In the following we make the explicit choice
(cf., e.g., \cite{39b}),
\begin{equation} \label{6.16}
\Gamma(E,x)=\begin{pmatrix}E-\tilde E-
\tfrac{i}{2}q(x)\phi^{(0)}(\tilde E,x)& \tfrac{i}{2}q(x)\\
\tfrac{i}{2}\phi^{(0)}(\tilde E,x)&-\tfrac{i}{2}\end{pmatrix},
\quad E\in \bbC \setminus \{\tilde E\}
\end{equation}
for some fixed $\tilde E \in \bbC$ and
\begin{equation} \label{6.19}
\phi^{(0)}(\tilde E,x)=\psi^{(0)}_2(\tilde E,x)/\psi^{(0)}_1(\tilde E,x),
\end{equation}
where $\Psi^{(0)}(\tilde E,x)=(\psi^{(0)}_1(\tilde E,x),
\psi^{(0)}_2(\tilde E,x))^t$ is any solution of $L\Psi=\tilde E\Psi$.
Using (\ref{6.7}), equation \eqref{6.14} becomes
\begin{align}
\tilde p(x) &= \phi^{(0)}(\tilde E,x), \label{6.17} \\
\tilde q(x) &= -q_x(x)-2i\tilde E q(x)+q(x)^2
\phi^{(0)}(\tilde E,x). \label{6.18}
\end{align}
Moreover, one computes for $\tilde \Psi = (\tilde \psi_1,\tilde
\psi_2)^t$ in terms of $\Psi=(\psi_1,\psi_2)^t$,
\begin{align}
\tilde \psi_1(E,x)&=(E-\tilde E)\psi_1(E,x) + \frac{i}{2}q(x)
(\psi_2(E,x)-\phi^{(0)}(\tilde E,x)\psi_1(E,x)), \label{6.20} \\
\tilde \psi_2(E,x)&= -\frac{i}{2}
(\psi_2(E,x)-\phi^{(0)}(\tilde E,x)\psi_1(E,x)).
\end{align}
In addition, we note that
\begin{equation} \label{6.21}
\det(\Gamma(E,x))=-\frac{i}{2}(E-\tilde E)
\end{equation}
and therefore,
\begin{equation} \label{6.22}
W(\tilde \Psi_1(E),\tilde \Psi_2(E))=-\frac{i}{2}(E-\tilde E)
W(\Psi_1(E),\Psi_2(E)),
\end{equation}
where $\Psi_j(E,x)$, $j=1,2$ are two linearly independent solutions of
$L\Psi=E\Psi$.

Our first result proves that gauge transformations as defined in this
section leave the class of meromorphic algebro-geometric potentials
of the AKNS hierarchy invariant.
\begin{thm} \label{t6.1}
Suppose $(p,q)$ is a meromorphic algebro-geometric AKNS potential. Fix
$\tilde E\in \bbC$ and define $(\tilde p,\tilde q)$ as in (\ref{6.17}),
(\ref{6.18}), with $\phi^{(0)}(\tilde E,x)$ defined as in (\ref{6.19}).
Suppose $\phi^{(0)}(\tilde E,x)$ is meromorphic in $x$. Then $(\tilde
p,\tilde q)$ is a meromorphic algebro-geometric AKNS potential.
\end{thm}

\begin{proof}
The upper right entry $G_{1,2}(E,x,x')$ of the Green's matrix of $L$ is
given by
\begin{equation}
G_{1,2}(E,x,x')=\frac{i\psi_{+,1}(E,x,x_0)\psi_{-,1}(E,x',x_0)}
{W(\Psi_-(E,\cdot,x_0),\Psi_+(E,\cdot,x_0))}, \quad x \geq x'.
\end{equation}
Combining (\ref{6.2})--(\ref{6.4}) and (\ref{6.7}), its diagonal (where
$x=x'$) equals
\begin{equation}\label{6.26}
G_{1,2}(E,x,x)=\frac{iF_n(E,x)}{2V_+(E)}.
\end{equation}
The corresponding diagonal of the upper right entry
$\tilde G_{1,2}(E,x,x)$ of the
Green's matrix of $\tilde L$ is computed to be
\begin{equation}\label{6.29}
\tilde G_{1,2}(E,x,x)=\frac{i\tilde\psi_{+,1}(E,x)\tilde\psi_{-,1}(E,x)}
{W(\tilde \Psi_-(E,\cdot),\tilde \Psi_+(E,\cdot))}
=\frac{i\tilde F_{n+1}(E,x)}{2(E-\tilde E)V_+(E)},
\end{equation}
where
\begin{align}\label{01052}
\tilde F_{n+1}(E,x)=&2iF_n(E,x) \{(E-\tilde E)
+\tfrac{i}{2}q(x)[\phi_+(E,x)-\phi^{(0)}(\tilde E,x)]\} \times \notag \\
&\times \{(E-\tilde E)+\tfrac{i}{2}q(x) [\phi_-(E,x)
-\phi^{(0)}(\tilde E,x)]\},
\end{align}
using \eqref{6.8}, \eqref{6.20}, \eqref{6.22}, and
\begin{equation} \label{6.30}
\psi_{+,1}(E,x,x_0)\psi_{-,1}(E,x,x_0)=\frac{F_n(E,x)}{F_n(E,x_0)}.
\end{equation}
By \eqref{6.2}, $\phi_+(E,x)+\phi_-(E,x)
=2G_{n+1}(E,x)/F_n(E,x)$ and $\phi_+(E,x)\phi_-(E,x)
=$ $H_n(E,x)/F_n(E,x)$. From this, \eqref{2.5}, and \eqref{2.13}, it
follows that $\tilde F_{n+1}(\cdot,x)$ is a polynomial of degree $n+1$
with leading coefficient
\begin{equation}\label{6.28a}
i q_x(x)-2\tilde E q(x)  -i q(x)^2 \phi^{(0)}(\tilde E,x)
=-i\tilde q(x).
\end{equation}

Finally, using $\tilde L\tilde\Psi_{\pm}=E \tilde \Psi_{\pm}$, that is,
(\ref{6.15}), one verifies that $\tilde G_{1,2}(E,x,x)$ satisfies the
differential equation,
\begin{equation} \label{6.33}
\tilde q\bigl(2\tilde G_{1,2}\tilde G_{1,2,xx}-\tilde G_{1,2,x}^2
+4(E^2-\tilde p\tilde q)\tilde G_{1,2}^2\bigr)
-\tilde q_x \bigl(2\tilde G_{1,2}\tilde G_{1,2,x}
+4iE\tilde G_{1,2}^2\bigr)=\tilde q^3.
\end{equation}
Hence $\tilde F_{n+1}(E,x)$ satisfies the hypotheses of Corollary
\ref{c2.3a} (with $n$ replaced by $n+1$ and $(p,q)$ replaced by $(\tilde
p,\tilde q)$) and therefore, $(\tilde p, \tilde q)$ is a meromorphic
algebro-geometric AKNS potential.
\end{proof}

We note here that \eqref{6.33} implies also that $\tilde F_{n+1}$
satisfies
\begin{align}
&\tilde q(2\tilde F_{n+1}\tilde F_{n+1,xx}-\tilde F_{n+1,x}^2
+4(E^2-\tilde p\tilde q)\tilde F_{n+1}^2)
-\tilde q_x(2\tilde F_{n+1}\tilde F_{n+1,x}+4iE \tilde F_{n+1}^2)
\notag\\
=&-4 \tilde q^3 (E-\tilde E)^2 R_{2n+2}(E), \label{02051}
\end{align}
that is, $(\tilde p,\tilde q)$ is associated with the curve
\begin{equation}\label{02052}
\tilde \calK_{n+1}=\{(E,V) \, | \, V^2=(E-\tilde E)^2 R_{2n+2}(E)\}.
\end{equation}

\begin{cor} \label{c6.2}
Suppose $(p,q)$ is a meromorphic algebro-geometric AKNS potential
associated with the hyperelliptic curve
\begin{equation}
\calK_n=\{(E,V) \, | \,V^2=R_{2n+2}(E)=\prod^{2n+1}_{m=0}(E-E_m)\},
\end{equation}
which has a singular point at $(\tilde E,0)$, that is, $R_{2n+2}$ has a
zero of order $r\geq2$ at the point $\tilde E$. Choose
\begin{equation} \label{6.35}
\phi^{(0)}(\tilde E,x)=\frac{G_{n+1}(\tilde E,x)}
{F_n(\tilde E,x)}
\end{equation}
(cf. (\ref{6.2})) and define $(\tilde p,\tilde q)$ as in
(\ref{6.17}), (\ref{6.18}). Then $\phi^{(0)}(\tilde E,\cdot)$ is
meromorphic and the meromorphic algebro-geometric AKNS potential
$(\tilde p,\tilde q)$ is associated with the hyperelliptic curve
\begin{equation}\label{6.36}
\tilde \calK_{\tilde n}=\{(E,V) \, | \,V^2=\tilde R_{2n-2s+4}(E)
=(E-\tilde E)^{2-2s} R_{2n+2}(E)\}
\end{equation}
for some $2\leq s\leq (r/2)+1$. In particular, $\tilde \calK_{\tilde n}$
and $\calK_n$ have the same structure near any point $E\neq \tilde E$.
\end{cor}

\begin{proof}
Since $V^2=R_{2n+2}(E)$ we infer that $V_+(E)$ has at least a simple
zero
at $\tilde E$. Hence
\begin{equation}\label{01051}
\phi_\pm(E,x)-\phi^{(0)}(\tilde E,x)=\frac{\pm V_+(E)}{F_n(E,x)}
+\frac{G_{n+1}(E,x)F_n(\tilde E,x)-G_{n+1}(\tilde E,x)F_n(E,x)}
{F_n(E,x)F_n(\tilde E,x)}
\end{equation}
also have at least a simple zero at $\tilde E$. From \eqref{01052} one
infers that $\tilde F_{n+1}(E,x)$ has a zero
of order at least $2$ at $\tilde E$, that is,
\begin{equation}\label{02053}
\tilde F_{n+1}(E,x)=(E-\tilde E)^s \tilde F_{n+1-s}(E,x), \,\, s\geq 2.
\end{equation}
Define $\tilde n=n+1-s$. Then $\tilde F_{\tilde n}$ still satisfies the
hypothesis of Corollary \ref{c2.3a}. Moreover, inserting \eqref{02053}
into \eqref{02051} shows that $(E-\tilde E)^{2s}$ must be a factor of
$(E-\tilde E)^2 R_{2n+2}(E)$. Thus, $2s\leq r+2$ and hence
\begin{equation}
\tilde q(2\tilde F_{\tilde n}\tilde F_{\tilde n,xx}
-\tilde F_{\tilde n,x}^2+4(E^2-\tilde p\tilde q)\tilde F_{\tilde n}^2)
-\tilde q_x(2\tilde F_{\tilde n}\tilde F_{\tilde n,x}
+4iE \tilde F_{\tilde n}^2)
=-4 \tilde q^3 \tilde R_{2\tilde n+2}(E), \label{02054}
\end{equation}
where
\begin{equation}\label{02055}
\tilde R_{2\tilde n+2}(E)=(E-\tilde E)^{2-2s} R_{2n+2}(E)
\end{equation}
is a polynomial in $E$ of degree $0<2n-2s+4<2n+2$. This proves
\eqref{6.36}.
\end{proof}

In view of our principal result, Theorem \ref{t5.4}, our choice of
$\phi^{(0)}(\tilde E,x)$ led to a curve $\tilde \calK_{\tilde n}$ which
is less singular at $\tilde E$ than $\calK_n$, without changing the
structure of the curve elsewhere. By iterating the procedure from
$\calK_n$ to $\tilde\calK_{\tilde n}$ one ends up with a curve which is
nonsingular at $(\tilde E,0)$. Repeating this procedure for each
singular point of $\calK_n$ then results in a nonsingular curve $\hat
\calK_{\hat n}$ and a corresponding Baker-Akhiezer function $\hat
\Psi(P,x,x_0)$ which is meromorphic with respect to $x\in \bbC$ (this
can be seen by using their standard theta function representation, cf.,
e.g., \cite{27b}). Suppose that $\hat \calK_{\hat n}$ was obtained from
$\calK_n$ by applying the gauge transformation
\begin{equation}
\Gamma(E,x)=\Gamma_N(E,x)...\Gamma_1(E,x),
\end{equation}
where each of the $\Gamma_j$ is of the type \eqref{6.16}. Then the
branches of
\begin{equation}\label{02056}
\Psi(P,x)=\Gamma(E,x)^{-1} \hat \Psi(P,x,x_0)
=\Gamma_1(E,x)^{-1}...\Gamma_N(E,x)^{-1} \hat \Psi(P,x,x_0)
\end{equation}
are linearly independent solutions of $L\Psi=E\Psi$ for all $E\in
\bbC \setminus \{E_0,...,E_{2n+1},\mu_1(x_0),...,\linebreak[0]
\mu_n(x_0)\}$.
These branches are meromorphic with respect to $x$ since
\begin{equation}\label{02057}
\Gamma_j(E,x)^{-1}=\frac{2i}{E-\tilde E}
\begin{pmatrix}-\tfrac{i}{2}&-\tfrac{i}{2}q(x)\\
-\tfrac{i}{2}\phi^{(0)}(\tilde E,x)
&E-\tilde E-\tfrac{i}{2}q(x)\phi^{(0)}(\tilde E,x)
\end{pmatrix}
\end{equation}
maps meromorphic functions to meromorphic functions in view of the fact
that $q$ and $\phi^{(0)}(\tilde E,\cdot)=G_{n+1}(\tilde E,\cdot)/
F_n(\tilde E,\cdot)$ are meromorphic. Combining these findings and
Theorem \ref{t05051} we thus proved the principal result of this section.
\begin{thm} \label{t6.3}
Suppose $(p,q)$ is a meromorphic algebro-geometric AKNS potential. Then
the solutions of $L\Psi=E\Psi$ are meromorphic with respect to the
independent variable for all values of the spectral parameter
$E\in \bbC$.
\end{thm}

\begin{rem}
In the
case of the KdV hierarchy, Ehlers and Kn\"orrer \cite{17b} used the
Miura transformation and algebro-geometric methods to prove results
of the type stated in Corollary \ref{c6.2}. An alternative approach in
the KdV context has recently been found by Ohmiya \cite{49c}. The
present technique to combine gauge transformations, the polynomial
recursion approach to integrable hierarchies based on hyperelliptic
curves (such as the KdV, AKNS, and Toda hierarchies), and the
fundamental meromorphic function $\phi(\cdot,x)$ on $\calK_n$ (cf.
(\ref{6.2})), yields a relatively straightforward and unified treatment,
further details of which will appear elsewhere. To the best of our
knowledge this is the first such approach for the AKNS hierarchy.

A systematic study of the construction used in Theorem \ref{t6.3} yields
explicit connections between the $\tau$-function associated with the
possibly singular curve $\calK_n$ and the Riemann theta function of the
nonsingular curve $\hat \calK_{\hat n}$. This seems to be of independent
interest and will be pursued elsewhere.
\end{rem}

\section{Floquet Theory} \label{floquet}
\setcounter{equation}{0}
Throughout this section we will assume the validity of the following
basic hypothesis.
\begin{hyp} \label{h3.1}
Suppose that $p, q \in L^1_{\text{loc}} (\bb R)$ are
complex-valued periodic functions of period $\Omega>0$
and that $L$ is a $2 \times 2$ matrix-valued
differential expression of the form
\begin{equation} \label{3.0}
L=J\frac{d}{dx}+Q,
\end{equation}
where
\begin{equation} \label{3.1}
J=\begin{pmatrix}i&0\\0&-i\end{pmatrix}, \quad
Q=\begin{pmatrix}0&-iq\\ip&0\end{pmatrix}.
\end{equation}
\end{hyp}
We note that
\begin{equation} \label{3.2}
-J^2=I \quad \text{ and } \quad JQ+QJ=0,
\end{equation}
where $I$ is the $2\times2$ identity matrix in $\bb C^2$.

Given Hypothesis \ref{h3.1}, we uniquely associate the following densely
and maximally defined closed linear operator $H$ in $L^2(\bb R)^2$ with
the matrix-valued differential expresssion $L$,
\begin{equation} \label{3.2a}
Hy=Ly, \, {\mathcal D}(H)=\{y \in L^2 (\bb R)^2 \, | \, y \in
AC_{\text{loc}}(\bb R)^2, \, Ly \in L^2(\bb R)^2\}.
\end{equation}
One easily verifies that $L$ is unitarily equivalent to
\begin{equation} \label{3.3}
\begin{pmatrix}0&-1\\1&0\end{pmatrix}\frac{d}{dx}+
\frac12
\begin{pmatrix}(p+q)&i(p-q)\\i(p-q)&-(p+q)\end{pmatrix},
\end{equation}
a form widely used in the literature.

We consider the differential equation $Ly=Ey$ where $L$ satisfies
Hypothesis \ref{h3.1} and where $E$ is a complex spectral parameter.
Define $\phi_0(E,x,x_0,Y_0)=\e^{E(x-x_0)J}Y_0$ for $Y_0\in M_2(\bb C)$.
The matrix function $\phi(E,\cdot,x_0,Y_0)$ is the unique solution of
the integral equation
\begin{equation} \label{3.4}
Y(x)=\phi_0(E,x,x_0,Y_0)+\int_{x_0}^x \e^{E(x-x')J}
JQ(x') Y(x') dx'
\end{equation}
if and only if it satisfies the initial value problem
\begin{equation} \label{3.5}
JY'+QY=EY, \quad Y(x_0)=Y_0.
\end{equation}

Since
\begin{equation} \label{3.6}
\frac{\partial \phi_0}{\partial x_0}(E,x,x_0,Y_0)
=E J \e^{E(x-x_0)J}Y_0=E \e^{E(x-x_0)J} JY_0,
\end{equation}
differentiating \eqref{3.4} with respect to $x_0$ yields
\begin{align}
\frac{\partial \phi}{\partial x_0}(E,x,x_0,Y_0)
=& \e^{E(x-x_0)J}(EJ-JQ(x_0))Y_0 \notag \\
&+\int_{x_0}^x \e^{E(x-x')J} J Q(x')
\frac{\partial \phi}{\partial x_0}(E,x',x_0,Y_0,) dx',
 \label{3.7}
\end{align}
that is,
\begin{equation} \label{3.8}
\frac{\partial \phi}{\partial x_0}(E,x,x_0,Y_0)
=\phi(E,x,x_0,(E+Q(x_0))JY_0),
\end{equation}
taking advantage of the fact that $\eqref{3.4}$ has unique solutions.

In contrast to the Sturm-Liouville case, the Volterra integral equation
\eqref{3.4} is not suitable to determine the asymptotic behavior of
solutions as $E$ tends to infinity. The following treatment circumvents
this difficulty and closely follows the outline in \cite{44e},
Section~1.4.

Suppose $L$ satifies Hypothesis \ref{h3.1}, $p,q \in C^n(\bb R)$, and
then define recursively,
\begin{align}
a_1(x)&=iQ(x), \notag \\
b_k(x)&=-i\int_0^x Q(t) a_k(t) dt, \label{3.9} \\
a_{k+1}(x)&=-a_{k,x}(x)+iQ(x) b_k(x), \quad k=1,
\dots,n.  \notag
\end{align}

Next let $A:\bb R^2\to \bb M_2(\bb C)$ be the unique solution of the
integral equation
\begin{equation} \label{3.10}
A(x,y)=a_{n+1}(x-y)+\int_0^y Q(x-y')
\int_{y-y'}^{x-y'} Q(x')
A(x',y-y') dx' dy'.
\end{equation}
Introducing
\begin{equation} \label{3.11}
\hat a_n(E,x)=\int_0^x A(x,y) \e^{-2iEy}dy, \quad
\hat b_n(E,x)=-i\int_0^x Q(y) \hat a_n(E,y) dy
\end{equation}
and
\begin{align}
u_1(E,x)&=I+\sum_{k=1}^n b_k(x) (2iE)^{-k}
+\hat b_n(E,x)(2iE)^{-n}, \label{3.12} \\
u_2(E,x)&=\sum_{k=1}^n a_k(x) (2iE)^{-k}
+\hat a_n(E,x)(2iE)^{-n}, \label{3.13}
\end{align}
we infer that
\begin{align}
Y_1(E,x)&=\e^{iEx} \{(I+iJ)u_1(E,x)+(I-iJ)u_2(E,x)\},
\label{3.14}\\
Y_2(E,x)&=\e^{-iEx} \{(I-iJ)u_1(-E,x)-(I+iJ)u_2(-E,x)\}
\label{3.15}
\end{align}
satisfy the differential equation
\begin{equation} \label{3.16}
JY'+QY=EY.
\end{equation}
Since $|A(x,y)|$ is bounded on compact subsets of $\bb R^2$ one obtains
the estimates
\begin{equation} \label{3.17}
|\e^{iEx}\hat a_n(E,x)|, \quad |\e^{iEx}
\hat b_n(E,x)|\leq CR^2 \e^{|x\Im(E)|}
\end{equation}
for a suitable constant $C>0$ as long as $|x|$ is bounded by some
$R>0$.

The matrix $\hat Y(E,x,x_0)=(Y_1(E,x-x_0)+Y_2(E,x-x_0))/2$ is also
a solution
of $JY'+QY=EY$ and satisfies $\hat Y(E,x_0,x_0)=I+Q(x_0)/(2E)$.
Therefore, at
least for sufficiently large $|E|$, the matrix function
\begin{equation} \label{3.18}
\phi(E,\cdot,x_0,I)= \hat Y(E,\cdot,x_0) \hat Y(E,x_0,x_0)^{-1}
\end{equation}
is the unique solution of the initial value problem $JY'+QY=EY$,
$Y(x_0)=I$. Hence, if $p, q \in C^2(\bb R)$, one obtains the asymptotic
expansion
\begin{align}
\phi(E,x_0 + \Omega,x_0,I)=&
\begin{pmatrix} \e^{-iE\Omega}&0\\0&\e^{iE\Omega}\end{pmatrix}
+\frac{1}{2iE}\begin{pmatrix}\beta\e^{-iE\Omega}&2q(x_0)\sin(E\Omega)\\
2p(x_0)\sin(E\Omega) &-\beta\e^{iE\Omega}\end{pmatrix}
\notag \\
&+O(\e^{|\Im(E)|\Omega}E^{-2}), \label{3.19}
\end{align}
where
\begin{equation}
\beta=\int_{x_0}^{x_0+\Omega} p(t)q(t) dt. \label{3.19a}
\end{equation}
 From this result we infer in particular that the entries of
$\phi(\cdot,x_0 +\Omega,x_0,I)$, which are entire functions, have order
one whenever $q(x_0)$ and $p(x_0)$ are nonzero.

Denote by $T$ the operator defined by $Ty=y(\cdot+\Omega)$ on the set of
$\bb C^2$-valued functions on $\bb R$ and suppose $L$ satisfies
Hypothesis \ref{h3.1}. Then $T$ and $L$ commute and this implies that
$T(E)$, the restriction of $T$ to the (two-dimensional) space $V(E)$ of
solutions of $Ly=Ey$, maps $V(E)$ into itself. Choosing as a basis of
$V(E)$ the columns of $\phi(E,\cdot,x_0,I)$, the operator $T(E)$ is
represented by the matrix $\phi(E,x_0 + \Omega,x_0,I)$. In particular,
$\det (T(E))=\det(\phi(E,x_0 +\Omega,x_0,I))=1$. Therefore, the
eigenvalues $\rho(E)$ of $T(E)$, the so called Floquet multipliers, are
determined as solutions of
\begin{equation} \label{3.20}
\rho^2 -\tr (T(E)) \rho +1=0.
\end{equation}
These eigenvalues are degenerate if and only if $\rho^2 (E)=1$ which
happens if and only if the equation $Ly=Ey$ has a solution of period
$2\Omega$. Hence we now study asymptotic properties of the spectrum of
the densely defined closed realization $H_{2\Omega,x_0}$ of $L$ in
$L^2([x_0,x_0 + 2\Omega])^2$ given by
\begin{align}
H_{2\Omega,x_0}y=Ly,  \, {\mathcal D} (H_{2\Omega,x_0}) =
\{& y \in L^2([x_0,x_0+2\Omega])^2 \, | \,
y\in AC([x_0,x_0 + 2\Omega])^2, \notag \\
& y(x_0+2\Omega)=y(x_0), \,
Ly \in L^2([x_0,x_0+2\Omega])^2 \}. \label{3.21}
\end{align}
Its eigenvalues, which are called the (semi-)periodic eigenvalues of
$L$, and their multiplicities are given, respectively, as the zeros and
their multiplicities of the function $\tr (T(E))^2 -4$. The asymptotic
behavior of these eigenvalues is described in the following result.
\begin{thm} \label{t3.2}
Suppose that $p,q \in C^2(\bb R)$. Then the eigenvalues $E_j$, $j\in\bb
Z$ of $H_{2\Omega,x_0}$ are $x_0$-independent and satisfy the asymptotic
behavior
\begin{equation} \label{3.22}
E_{2j},E_{2j-1}=\frac{j\pi}{\Omega}+O(\frac1{|j|})
\end{equation}
as $|j|$ tends to infinity, where all eigenvalues are repeated according
to their algebraic multiplicities. In particular, all eigenvalues of
$H_{2\Omega,x_0}$ are contained in a strip
\begin{equation} \label{3.23}
\Sigma=\{E\in\bb C \, | \, |\Im(E)|\leq C\}
\end{equation}
for some constant $C>0$.
\end{thm}
\begin{proof}
Denoting $A(E,x)=(E+Q(x))J$ (cf.(\ref{6.11})), equation \eqref{3.8}
implies
\begin{align}
&\partial \phi(E,x, x_0,I)/\partial x
=-A(E,x)\phi(E,x,x_0,I), \label{3.31} \\
&\partial \phi(E,x,x_0,I)/\partial x_0
=\phi(E,x,x_0,I)A(E,x) \label{3.32}
\end{align}
and hence
\begin{equation} \label{3.24a}
\partial \tr(T(E))/\partial x_0 =0.
\end{equation}
Thus the eigenvalues of $H_{2\Omega,x_0}$ are independent of $x_0$.
According to \eqref{3.19}, $\tr (T(E))$ is asymptotically given by
\begin{equation} \label{3.24}
\tr (T(E))=2\cos (E\Omega)+\beta \sin(E\Omega)E^{-1}
+ O(\e^{|\Im(E)|\Omega}E^{-2}).
\end{equation}
Rouch\'e's theorem then implies that two eigenvalues $E$ lie in a
circle
centered at $j\pi/a$ with radius of order $1/|j|$.

To prove that the eigenvalues may be labeled in the manner indicated,
one again uses Rouch\'e's theorem with a circle of sufficiently large
radius centered at the origin of the $E$-plane in order to compare the
number of zeros of $\tr (T(E))^2 -4$ and $4\cos(E\Omega)^2 -4$ in the
interior of this circle.
\end{proof}

The conditional stability set $\mc S(L)$ of $L$ in \eqref{3.0} is
defined to be the set of all spectral parameters $E$ such that $Ly=Ey$
has at least one bounded nonzero solution. This happens if and only if
the Floquet multipliers $\rho (E)$ of $Ly=Ey$ have absolute value one.
Hence
\begin{equation} \label{3.25}
\mc S(L)=\{E\in\bb C \, | \, -2 \leq \tr (T(E))\leq 2\}.
\end{equation}
It is possible to prove that the spectrum of $H$ coincides with the
conditional stability set $\mc S(L)$ of $L$, but since we do not need
this fact we omit a proof. In the following we record a few
properties of $\mc S(L)$ to be used in Sections \ref{fingap} and
\ref{picard}.
\begin{thm} \label{t3.3}
Assume that $p,q \in C^2(\bb R)$. Then the conditional stability set
$\mc S(L)$ consists of a countable number of regular analytic arcs, the
so called spectral bands. At most two spectral bands extend to infinity
and at most finitely many spectral bands are closed arcs. The point $E$
is a band edge, that is, an endpoint of a spectral band, if and only if
$\tr (T(E))^2 -4$ has a zero of odd order.
\end{thm}
\begin{proof}
The fact that $\mc S(L)$ is a set of regular analytic arcs whose
endpoints are odd order zeros of $(\tr (T(E)))^2 -4$ and hence countable
in number, follows in standard manner from the fact that $\tr (T(E))$ is
entire with respect to $E$. (For additional details on this problem,
see, for instance, the first part of the proof of Theorem 4.2 in
\cite{w3}.)

 From the asymptotic expansion \eqref{3.19} one infers that $\tr (T(E))$
is approximately equal to $2\cos(E\Omega)$ for $|E|$ sufficiently large.
This implies that the Floquet multipliers are in a neighborhood of
$\e^{\pm iE\Omega}$. If $E_0\in\mc S(L)$ and $|E_0|$ is sufficiently
large, then it is close to a real number. Now let $E=|E_0|\e^{it}$,
where $t\in(-\pi/2,3\pi/2]$. Whenever this circle intersects $\mc S(L)$
then $t$ is close to $0$ or $\pi$. When $t$ is close to $0$, the Floquet
multiplier which is near $\e^{iE\Omega}$ moves radially inside the unit
circle while the one close to $\e^{-iE\Omega}$ leaves the unit disk at
the same time. Since this can happen at most once, there is at most one
intersection of the circle of radius $|E_0|$ with $\mc S(L)$ in the
right half-plane for $|E|$ sufficiently large. Another such
intersection may
take place in the left half-plane. Hence at most two arcs extend to
infinity and there are no closed arcs outside a sufficiently large disk
centered at the origin.

Since there are only countably many endpoints of spectral arcs, and
since outside a large disk there can be no closed spectral arcs, and at
most two arcs extend to infinity, the conditional stability set consists
of at most countably many arcs.
\end{proof}

Subsequently we need to refer to components of vectors in $\bb C^2$. If
$y\in\bb C^2$, we will denote the first and second components of $y$ by
$y_1$ and $y_2$, respectively, that is, $y=(y_1,y_2)^t$, where the
superscript ``$t$'' denotes the transpose of a vector in $\bb C^2$.

The boundary value problem $Ly=zy$, $y_1(x_0)=y_1(x_0+\Omega)=0$ in
close analogy to the scalar Sturm-Liouville case, will be called the
Dirichlet problem for the interval $[x_0,x_0+\Omega]$ and its
eigenvalues will therefore be called Dirichlet eigenvalues (associated
with the interval $[x_0,x_0+\Omega]$). In the corresponding operator
theoretic formulation one introduces the following closed realization
$H_{D,x_0}$ of $L$ in $L^2([x_0,x_0+\Omega])^2$,
\begin{align}
H_{D,x_0} y=Ly, \, {\mathcal D} (H_{D,x_0}) =
\{& y \in L^2([x_0,x_0+\Omega])^2 \, | \,
y\in AC([x_0,x_0 + \Omega])^2, \notag \\ & y_1(x_0)=y_1(x_0+\Omega)=0,
\, Ly \in L^2([x_0,x_0+\Omega])^2 \}. \label{3.26}
\end{align}
The eigenvalues of $H_{D,x_0}$ and their algebraic multiplicities are
given as the zeros and their multiplicities of the function
\begin{equation} \label{3.31a}
g(E,x_0)=(1,0)\phi(E,x_0+\Omega,x_0,I)(0,1)^t,
\end{equation}
that is, the entry in the upper right corner of
$\phi(E,x_0+\Omega,x_0,I)$.

\begin{thm} \label{t3.4}
Suppose $p,q \in C^2(\bb R)$. If $q(x_0)\neq0$ then there are countably
many Dirichlet eigenvalues $\mu_j(x_0)$, $j\in\bb Z$, associated with
the interval $[x_0,x_0+\Omega]$. These eigenvalues have the asymptotic
behavior
\begin{equation} \label{3.27}
\mu_j(x_0)=\frac{j\pi}{\Omega}+O(\frac1{|j|})
\end{equation}
as $|j|$ tends to infinity, where all eigenvalues are repeated
according
to their algebraic multiplicities.
\end{thm}

\begin{proof}
 From the asymptotic expansion \eqref{3.19} we obtain that
\begin{equation} \label{3.28}
g(E,x_0)=\frac{-iq(x_0)}{E}\sin(E\Omega)+O(\e^{|\Im(E)|\Omega}E^{-2}).
\end{equation}
Rouch\'e's theorem implies that one eigenvalue $E$ lies in a circle
centered at $j\pi/ \Omega$ with radius of order $1/|j|$ and that the
eigenvalues may be labeled in the manner indicated (cf. the proof of
Theorem \ref{t3.2}).
\end{proof}

We now turn to the $x$-dependence of the function $g(E,x)$.
\begin{thm} \label{t3.5}
Assume that $p,q \in C^1(\bb R)$. Then the function $g(E,\cdot)$
satisfies the differential equation
\begin{align}
&q(x)(2g(E,x) g_{xx}(E,x) -g_x(E,x)^2+4(E^2-
p(x)q(x))g(E,x)^2) \notag \\
&-q_x(x)(2g(E,x) g_x(E,x)+4iE g(E,x)^2)
=-q(x)^3(\tr (T(E))^2 -4). \label{3.30}
\end{align}
\end{thm}

\begin{proof}
Since $g(E,x)=(1,0)\phi(E,x,x+\Omega,I) (0,1)^t$ we obtain from
\eqref{3.31} and \eqref{3.32},
\begin{align}
g_x(E,x)=&(1,0)(\phi(E,x,x+\Omega,I) A(E,x)-A(E,x)
\phi(E,x,x+\Omega,I))(0,1)^t, \label{3.33} \\
g_{xx}(E,x)=&(1,0)(\phi(E,x,x+\Omega,I) A(E,x)^2-
2A(E,x)\phi(E,x,x+\Omega,I) A(E,x) \notag \\
&+A(E,x)^2\phi(E,x,x+\Omega,I)
+\phi(E,x,x+\Omega,I) A_x(E,x) \notag \\
&-A_x(E,x)\phi(E,x,x+\Omega,I))(0,1)^t,
\label{3.34}
\end{align}
where we used periodicity of $A$, that is, $A(E,x+\Omega)=A(E,x)$.
This
yields the desired result upon observing that
$\tr(\phi(z,x+\Omega,x,I))=\tr (T(E))$ is independent of $x$.
\end{proof}

\begin{defn} \label{d3.6}
The algebraic multiplictiy of $E$ as a Dirichlet eigenvalue
$\mu(x)$ of
$H_{D,x}$ is denoted by $\delta(E,x)$. The quantities
\begin{equation} \label{3.35}
\delta_i(E)=\min\{\delta(E,x)\, | \,x\in\bb R\},
\end{equation}
and
\begin{equation} \label{3.36}
\delta_m(E,x)=\delta(E,x)-\delta_i(E)
\end{equation}
will be called the immovable part and the movable part of the
algebraic
multiplicity $\delta(E,x)$, respectively. The sum $\sum_{E\in\bb C}
\delta_m(E,x)$ is called the number of movable Dirichlet eigenvalues.
\end{defn}

If $q(x)\neq0$ the function $g(\cdot,x)$ is an entire function with
order of growth equal to one. The Hadamard factorization theorem
then implies
\begin{equation} \label{3.37}
g(E,x)=F_D(E,x) D(E),
\end{equation}
where
\begin{align}
F_D(E,x)&=g_D(x) \e^{h_D(x)E} E^{\delta_m(0,x)}
\prod_{\lambda\in {\bb C} \setminus \{0\}}
(1-(E/ \lambda))^{\delta_m(z,x)}
\e^{\delta_m(z,x)E}, \label{3.38} \\
D(E)&=\e^{d_0 E} E^{\delta_i(E)}\prod_{\lambda\in
{\bb C} \setminus \{0\}}
(1-(E/ \lambda))^{\delta_i(E)}
\e^{\delta_i(E)E},\label{3.39}
\end{align}
for suitable numbers $g_D(x)$ and $h_D(x)$ and $d_0$.

Define
\begin{equation} \label{3.39a}
U(E)=(\tr (T(E))^2 -4)/D(E)^2.
\end{equation}
Then Theorem \ref{t3.2} shows that
\begin{align}
 &-q(x)^3 U(E)\notag\\
 =&q(x)(2F_D(E,x)F_{D,xx}(E,x)- F_{D,x}(E,x)^2+
 4(E^2-q(x)p(x))F_D(E,x)^2)\notag \\
 &-q_x(x)(2F_D(E,x)F_{D,x}(E,x_0)+4iE F_D(E,x)^2).
\label{3.40}
\end{align}
As a function of $E$ the left-hand side of this equation is entire (see
Proposition 5.2 in \cite{w3} for an argument in a similar case).
Introducing $s(E)=\ord_E(\tr (T(E))^2 -4)$ we obtain the following
important result.
\begin{thm} \label{t3.7}
Under the hypotheses of Theorem \ref{t3.5}, $s(E)-2\delta_i(E)\geq0$ for
every $E\in\bb C$.
\end{thm}

We now define the sets ${\calE}_1=\{E\in\bb C \, | \, s(E)>0, \,
\delta_i(E)=0\}$ and ${\calE}_2=\{E\in\bb C \, | \,
s(E)-2\delta_i(E)>0\}$. Of course, $\calE_1$ is a subset of $\calE_2$
which, in turn, is a subset of the set of zeros of $\tr (T(E))^2-4$ and
hence isolated and countable.

\begin{thm} \label{t3.8}
Assume Hypothesis \ref{h3.1} and that $Ly=Ey$ has degenerate Floquet
multipliers $\rho$ (equal to $\pm1$) but two linearly independent
Floquet solutions. Then $E$ is an immovable Dirichlet eigenvalue, that
is, $\delta_i(E)>0$. Moreover, $\calE_1$ is contained in the set of all
those values of $E$ such that $Ly=Ey$ does not have two linearly
independent Floquet solutions.
\end{thm}

\begin{proof}
If $Ly=Ey$ has degenerate Floquet multipliers $\rho (E)$ but two
linearly independent Floquet solutions then every solution of
$Ly=Ey$ is
Floquet with multiplier $\rho (E)$. This is true, in particular,
for the
unique solution $y$ of the initial value problem $Ly=Ey$,
$y(x_0)=(0,1)^t$. Hence $y(x_0+\Omega)=(0,\rho)^t$ and $y$ is a
Dirichlet eigenfunction regardless of $x_0$, that is, $\delta_i(E)>0$.

If $E\in \calE_1$ then $s(E)>0$ and $Ly=Ey$ has degenerate Floquet
multipliers. Since $\delta_i(E)=0$, there cannot be two linearly
independent Floquet solutions.
\end{proof}

Spectral theory for nonself-adjoint periodic Dirac operators has
very recently drawn considerable attention in the literature and
we refer the reader to \cite{32e} and \cite{59c}.

\section{Floquet Theory and Algebro-Geometric Potentials} \label{fingap}
\setcounter{equation}{0}
In this section we will obtain necessary and sufficient conditions in
terms of Floquet theory for a function $Q:\bb R\to M_2(\bb C)$ which is
periodic with period $\Omega>0$ and which has zero diagonal entries to
be algebro-geometric (cf. Definition \ref{d2.2}). Throughout this
section we assume the validity of Hypothesis \ref{h3.1}.

We begin with sufficient conditions on $Q$ and recall the definition of
$U(E)$ in \eqref{3.39a}.

\begin{thm} \label{t4.1}
Suppose that $p,q \in C^2(\bb R)$ are periodic with period $\Omega>0$.
If $U(E)$ is a polynomial of degree $2n+2$ then the following
statements
hold. \\
 (i) $\deg(U)$ is even, that is, $n$ is an integer. \\
 (ii)The number of movable Dirichlet eigenvalues (counting algebraic
 multiplicities) equals $n$. \\
 (iii) $\mc S(L)$ consists of finitely many regular analytic arcs. \\
 (iv) $p,q\in C^\infty(\bb R)$. \\
 (v) There exists a $2\times2$ matrix-valued differential expression
 $P_{n+1}$ of order $n+1$ with leading coefficient $J^{n+2}$ which
 commutes with $L$ and satisfies
\begin{equation} \label{4.1}
P_{n+1}^2=\prod_{E\in F_2} (L-E)^{s(E)-2\delta_i(E)}.
\end{equation}
\end{thm}
\begin{proof}
The asymptotic behavior of Dirichlet and periodic eigenvalues (Theorems
\ref{t3.2} and \ref{t3.4}) shows that $s(E)\leq 2$ and $\delta(E,x)
\leq1$ when $|E|$ is suitably large. Since $(U(E)$ is a polynomial,
$s(E)>0$ implies that $s(E)=2\delta_i(E)=2$. If $\mu(x)$ is a Dirichlet
eigenvalue outside a sufficiently large disk, then it must be close to
$m\pi/ \Omega$ for some integer $m$ and hence close to a point $E$ where
$s(E)=2\delta_i(E)=2$. Since there is only one Dirichlet eigenvalue in
this vicinity we conclude that $\mu(x)=E$ is independent of $x$. Hence,
outside a sufficiently large disk, there is no movable Dirichlet
eigenvalue, that is, $F_D(\cdot,x)$ is a polynomial. Denote its degree,
the number of movable Dirichlet eigenvalues, by $\tilde n$. By
\eqref{3.40} $U(E)$ is a polynomial of degree $2\tilde n+2$. Hence
$\tilde n=n$ and this proves parts (i) and (ii) of the theorem.

Since asymptotically $s(E)=2$, we infer that $s(E)=1$ or $s(E)\geq 3$
occurs at only finitely many points $E$. Hence, by Theorem \ref{t3.3},
there are only finitely many band edges, that is, $\mc S(L)$ consists of
finitely many arcs, which is part (iii) of the theorem.

Let $\gamma(x)$ be the leading coefficient of $F_D(\cdot,x)$. From
equation \eqref{3.40} we infer that $-\gamma(x)^2/q(x)^2$ is the leading
coefficient of $U(E)$ and hence $\gamma(x)=ci q(x)$ for a suitable
constant $c$. Therefore, $F(\cdot,x)=F_D(\cdot,x)/c$ is a polynomial of
degree $n$ with leading coefficient $iq(x)$ satisfying the hypotheses of
Theorem \ref{t2.3}. This proves that $p, q\in C^\infty(\bb R)$ and that
there exists a $2 \times 2$ matrix-valued differential expression
$P_{n+1}$ of order $n+1$ and leading coefficient $J^{n+2}$ which
commutes with $L$. The differential expressions $P_{n+1}$ and $L$
satisfy $P_{n+1}^2=R_{2n+2}(L)$, where
\begin{equation} \label{4.2}
R_{2n+2}(E)=U(E)/(4c^2)=\prod_{\lambda\in F_2}
(E-\lambda)^{s(\lambda)-2\delta_i(\lambda)},
\end{equation}
concluding parts (iv) and (v) of the theorem.
\end{proof}

\begin{thm} \label{t4.2}
Suppose that $p, q \in C^2(\bb R)$ are periodic of period $\Omega > 0$
and that the differential equation $Ly=Ey$ has two linearly independent
Floquet solutions for all but finitely many values of $E$. Then $U(E)$
is a polynomial.
\end{thm}
\begin{proof}
Assume that $U(E)$ in \eqref{3.39a} is not a polynomial. At any point
outside a large disk where $s(E)>0$ we have two linearly independent
Floquet solutions and hence, by Theorem \ref{t3.8}, $\delta_i(E)\geq1$.
On the other hand, we infer from Theorem \ref{t3.2} that $s(E)\leq 2$
and hence $s(E)-2\delta_i(E)=0$. Therefore, $s(E)-2\delta_i(E)>0$
happens only at finitely many points and this contradiction proves that
$U(E)$ is a polynomial.
\end{proof}

\begin{thm} \label{t4.3}
Suppose that $p, q \in C^2(\bb R)$  are periodic of period $\Omega > 0$
and that the associated Dirichlet problem has $n$ movable eigenvalues
for some $n \in \bb N$. Then $U(E)$ is a polynomial of degree $2n+2$.
\end{thm}
\begin{proof}
If there are $n$ movable Dirichlet eigenvalues, that is, if
$\deg(F_D(\cdot,x))=n$ then \eqref{3.40} shows that $U(E)=(\tr (T(E))^2
-4)/D(E)^2$ is a polynomial of degree $2n+2$.
\end{proof}

Next we prove that $U(E)$ being a polynomial, or the number of movable
Dirichlet eigenvalues being finite, is also a necessary condition for
$Q$ to be algebro-geometric.
\begin{thm} \label{t4.4}
Suppose $L$ satisfies Hypothesis \ref{h3.1}. Assume there exists a $2
\times 2$ matrix-valued differential expression $P_{n+1}$ of order $n+1$
with leading coefficient $J^{n+2}$ which commutes with $L$ but that
there is no such differential expression of smaller order commuting with
$L$. Then $U(E)$ is a polynomial of degree $2n+2$.
\end{thm}

\begin{proof}
Without loss of generality we may assume that $P_{n+1}
=\hat P_{c_1,...,c_{n+1}}$ for suitable constants $c_j$. According to
the results in Section \ref{akns}, the polynomial
\begin{equation} \label{4.3}
F_n(E,x) =\sum_{\ell=0}^n f_{n-\ell}(c_1,...,c_{n-\ell})(x)E^{\ell}
\end{equation}
satisfies the hypotheses of Theorem \ref{t2.3}. Hence the coefficients
$f_{\ell}$ and the functions $p$ and $q$ are in $C^\infty(\bb R)$. Also
the $f_{\ell}$, and hence $P_{n+1}$, are periodic with period $\Omega$.

Next, let $\mu(x_0)$ be a movable Dirichlet eigenvalue. Since $\mu(x)$
is a continuous function of $x\in \bb R$ and since it is not constant,
there exists an $x_0 \in \bb R$ such that $s(\mu(x_0))=0$, that is,
$\mu(x_0)$ is neither a periodic nor a semi-periodic eigenvalue.
Suppose
that for this choice of $x_0$ the eigenvalue $\mu:=\mu(x_0)$ has
algebraic multiplicity $k$. Let $V=\ker ((H_{D,x_0}-\mu)^k)$ be the
algebraic eigenspace of $\mu$. Then $V$ has a basis $\{y_1,...,y_k\}$
such that $(H_{D,x_0}-\mu)y_j=y_{j-1}$ for $j=1,...,k$, agreeing that
$y_0=0$. Moreover, we introduce $V_m :=
\text{span} \, \{y_1,...,y_m\}$ and $V_0=\{0\}$. First we show by
induction that there exists a number $\nu$ such that $(T-\rho)y,
(P_{n+1}-\nu)y \in V_{m-1}$, whenever $y\in V_m$.

Let $m=1$. Then $(H_{D,x_0}-\mu)y=0$ implies $y=\alpha y_1$ for some
constant $\alpha$ and hence $y$ is a Floquet solution with multiplier
$\rho=y_{1,2}(x_0+\Omega)$, that is, $(T-\rho)y=0$. (We define, in
obvious notation, $y_{j,k}$, $k=1,2$ to be the $k$-th component of
$y_j$, $1\leq j \leq m$.) Since $P_{n+1}$ commutes with both $L$ and
$T$, we find that $P_{n+1}y$ is also a Floquet solution with multiplier
$\rho$. Since $s(\mu)=0$, the geometric eigenspace of $\rho$ is
one-dimensional and hence $P_{n+1}y=\nu y$ for a suitable constant
$\nu$.

Now assume that the statement is true for $1\leq m<k$. Let
$y=\sum_{j=1}^{m+1}\alpha_jy_j \in V_{m+1}$. Note that $(T-\rho)y$
satisfies Dirichlet boundary conditions. Hence
\begin{equation} \label{4.4}
(H_{D,x_0}-\mu)(T-\rho)y=(T-\rho)\sum_{j=1}^{m+1}
\alpha_j (H_{D,x_0}-\mu)y_j
=\sum_{j=1}^{m+1}\alpha_j (T-\rho)y_{j-1}
\end{equation}
is an element of $V_{m-1}$, say equal to $v=\sum_{j=1}^{m-1}\beta_j
y_j$. The nonhomogeneous equation $(H_{D,x_0}-\mu)w= v$ has the general
solution
\begin{equation} \label{4.5}
w=\sum_{j=1}^{m-1}\beta_j y_{j+1} + \alpha y_1,
\end{equation}
where $\alpha$ is an arbitary constant. Since $w$ is in $V_m$, the
particular solution $(T-\rho)y$ of $(H_{D,x_0}-\mu)w=v$ is in $V_m$
too.

Since $Q$ is infinitely often differentiable, so are the functions
$y_1,...,y_k$. Hence $L$ can be applied to $(P_{n+1}-\nu)y$ and one
obtains
\begin{equation} \label{4.6}
(L-\mu)(P_{n+1}-\nu)y=\sum_{j=1}^{m+1}
\alpha_j(P_{n+1}-\nu)y_{j-1}
=\sum_{j=1}^{m-1}\gamma_jy_j
\end{equation}
for suitable constants $\gamma_j$. Thus there are numbers $\alpha_1$
and
$\alpha_2$ such that
\begin{equation} \label{4.7}
(P_{n+1}-\nu)y=\sum_{j=1}^{m-1}\gamma_j y_{j+1}+
\alpha_1 \hat y+\alpha_2y_1,
\end{equation}
where $\hat y$ is the solution of $Ly=\mu y$ with $\hat y(x_0)=(1,0)^t$.
Note that $(P_{n+1} y-\nu y)_1(x_0)=(P_{n+1} y)_1(x_0)=\alpha_1$. Let
$w=(T-\rho)y$ and $v=(P_{n+1}-\nu)w$. Then $w\in V_m$ and $v\in
V_{m-1}$. Hence,
\begin{align}
(P_{n+1}y-\nu y)_1(x_0+a)&=(T(P_{n+1}-\nu)y)_1(x_0)
=((P_{n+1}-\nu)Ty)_1(x_0) \notag \\
&=(P_{n+1}(\rho y+w))_1(x_0) \notag\\ &=\rho (P_{n+1}y)_1(x_0)
+(P_{n+1}w)_1(x_0)=\rho \alpha_1. \label{4.8}
\end{align}
On the other hand, $(P_{n+1}y-\nu y)_1(x_0+a)=\alpha_1/\rho$ since
$\hat{y}_1(x_0+a)=1/\rho$. Thus $0=\alpha_1(\rho-1/\rho)$ which implies
$\alpha_1=0$ and $(P_{n+1}-\nu)y\in V_m$.

Hence we have shown that $T$ and $P_{n+1}$ map $V$ into itself. In
particular, $(P_{n+1}y)_1(x_0)$ = $0$ for every $y\in V$.

Next observe that the functions $y_1,...,y_k$ defined above satisfy
$(L-\mu)^j y_m=y_{m-j}$, agreeing that $y_m=0$ whenever $m\leq0$.
Consequently,
\begin{equation} \label{4.9}
L^jy_m=\sum_{r=0}^j \binom jr \mu^r (L-\mu)^{j-r}y_m
=\sum_{r=0}^j \binom jr \mu^r y_{m+r-j}.
\end{equation}
Moreover,
\begin{align}
P_{n+1}y_m&=-\sum_{j=0}^{n+1}\left(g_{n+1-j}J+ iA_{n-j}\right)L^jy_m
\notag \\ &=-\sum_{j=0}^{n+1}\sum_{r=0}^j \binom{j}{r}\mu^r
\begin{pmatrix}ig_{n+1-j}y_{m+r-j,1}-if_{n-j}y_{m+r-j,2} \\
-ig_{n+1-j}y_{m+r-j,2}+ih_{n-j}y_{m+r-j,1}\label{4.10}
\end{pmatrix}.
\end{align}
Since $(P_{n+1}y_j)_1(x_0)=y_{j,1}(x_0)=0$, evaluating the first
component of \eqref{4.10} at $x_0$ yields
\begin{align}
0&=(P_{n+1}y_m)_1(x_0)=i\sum_{\ell=0}^{n+1} y_{m-\ell,2}(x_0)
\frac1{\ell!}\sum_{j=\ell}^{n+1} j...(j-\ell+1)
\mu^{j-\ell}f_{n-j}(x_0) \notag \\
&=i\sum_{\ell=0}^n y_{m-\ell,2}(x_0)\frac1{\ell!}
\frac{\partial^\ell F_n}{\partial E^\ell}(\mu,x_0).
\label{4.11}
\end{align}
Letting $m$ run from 1 through $k$ shows that $\mu$ is a zero of
$F_n(\cdot,x_0)$ of order at least $k$. Therefore, there can be at most
$n$ movable Dirichlet eigenvalues counting multiplicities. However, if
there were less than $n$ movable Dirichlet eigenvalues then, by Theorems
\ref{t4.1} and \ref{t4.3}, there would exist a differential expression
of order less than $n+1$ which commutes with $L$ without being a
polynomial of $L$. Hence there are precisely $n$ movable Dirichlet
eigenvalues and $\text{deg}\,(U)=2n+2$.
\end{proof}

\section{A Characterization of Elliptic Algebro-Geometric AKNS
Potentials} \label{picard}
\setcounter{equation}{0}
Picard's theorem yields sufficient conditions for a linear (scalar)
$n^{\rm th}$-order differential equation, whose coefficients are
elliptic functions with a common period lattice spanned by $2\omega_1$
and $2\omega_3$, to have a fundamental system of solutions which are
elliptic of the second kind. We start by generalizing Picard's theorem
to first-order systems. Let $T_j$, $j=1,3$, be the operators defined by
$T_jy=y(\cdot+2\omega_j)$. In analogy to the scalar case we call $y$
elliptic of the second kind if it is meromorphic and
\begin{equation} \label{5.0}
y(\cdot + 2\omega_j) = \rho_j y(\cdot) \, \text{for
some} \, \rho_j \in
\bb C \setminus \{0\}, \, j=1,3.
\end{equation}

\begin{thm} \label{t5.1}
Suppose that the entries of $A:\bb C \to M_n(\bb C_\infty)$ are elliptic
functions with common fundamental periods $2\omega_1$ and $2\omega_3$.
Assume that the first-order differential system $\psi'=A\psi$ has a
meromorphic fundamental system of solutions. Then there exists at least
one solution $\psi_1$ which is elliptic of the second kind. If in
addition, the restriction of either $T_1$ or $T_3$ to the
($n$-dimensional) space $W$ of solutions of $\psi'=A\psi$ has distinct
eigenvalues, then there exists a fundamental system of solutions of
$\psi'=A\psi$ which are elliptic of the second kind.
\end{thm}

\begin{proof}
$T_1$ is a linear operator mapping $W$ into itself and thus has an
eigenvalue $\rho_1$ and an associated eigenfunction $u_1$, that is,
$\psi'=A\psi$ has a solution $u_1$ satisfying $u_1(x+2\omega_1)
= \rho_1 u_1(x)$.

Now consider the functions
\begin{equation} \label{5.1}
u_1(x), \, u_2(x)=u_1(x+2\omega_3),\,..., \,
u_m(x)=u_1(x+2(m-1)\omega_3),
\end{equation}
where $m\in\{1,...,n\}$ is chosen such that the functions in
\eqref{5.1} are linearly independent but including
$u_1(x+2m\omega_3)$ would render a linearly dependent set of functions.
Then,
\begin{equation} \label{5.2}
u_m(x+2\omega_3)=b_1 u_1(x)+....+ b_m u_m(x).
\end{equation}
Next, denote the restriction of $T_3$ to the span $V$ of
$\{u_1,...,u_m\}$ by ${\tilde T_3}$. It follows from \eqref{5.2} that
the range of ${\tilde T_3}$ is again $V$. Let ${\rho_3}$ be an
eigenvalue of ${\tilde T_3}$ and $v$ the associated eigenvector, that
is, $v$ is a meromorphic solution of the differential equation
$\psi'=A\psi$ satisfying $v(x+2\omega_3)={\rho_3} v(x)$. But $v$ also
satisfies $v(x+2\omega_1)=\rho_1 v(x)$ since every element of $V$ has
this property. Hence $v$ is elliptic of the second kind.

The numbers $\rho_1$ and ${\rho_3}$ are the Floquet multipliers
corresponding to the periods $2\omega_1$ and $2\omega_3$, respectively.
The process described above can be performed for each multiplier
corresponding to the period $2\omega_1$. Moreover, the roles of
$2\omega_1$ and $2\omega_3$ may of course be interchanged. The last
statement of the theorem follows then from the observation that
solutions associated with different multipliers are linearly
independent.
\end{proof}

What we call Picard's theorem following the usual convention in
\cite{3}, p. 182--185, \cite{13}, p. 338--343, \cite{34}, p. 536--539,
\cite{41}, p. 181--189, appears, however, to have a longer history. In
fact, Picard's investigations \cite{53}--\cite{55} in the scalar
$n^{\rm
th}$-order case were inspired by earlier work of Hermite in the special
case of Lam\'e's equation \cite{35}, p. 118--122, 266--418, 475--478
(see also \cite{6a}, Sect. 3.6.4 and \cite{67}, p. 570--576). Further
contributions were made by Mittag-Leffler \cite{46}, and Floquet
\cite{21}--\cite{23}. Detailed accounts on Picard's differential
equation can be found in \cite{34}, p. 532--574, \cite{41},
p. 198--288.
For a recent extension of Theorem \ref{t5.1} see \cite{27a}.

Picard's Theorem \ref{t5.1} motivates the following definition.
\begin{defn} \label{d5.2}
A $2 \times 2$ matrix $Q$ whose diagonal entries are zero and whose
off-diagonal entries are elliptic functions with a common period
lattice
is called a {\bf Picard-AKNS potential} if and only if the differential
equation $J\psi'+Q\psi=E\psi$ has a meromorphic fundamental system of
solutions (with respect to the independent variable) for infinitely
many
values of the spectral parameter $E\in\bb C$.
\end{defn}
Recall from Theorem \ref{t05051} that $J\psi'+Q\psi=E\psi$ has a
meromorphic fundamental system of solutions for all values of $E$ if
this is true for a sufficiently large finite number of values of $E$.

In the following assume, without loss of generality, that
$\Re(\omega_1)>0$, $\Re(\omega_3) \ge 0$, $\Im(\omega_3/\omega_1)>0$.
The fundamental period parallelogram then consists of the points
$E=2\omega_1 s +2\omega_3 t$, where $0\leq s,t<1$.

We introduce $\theta\in(0,\pi)$ by
\begin{equation} \label{5.3}
e^{i\theta} = \frac{\omega_3}{\omega_1}
\left|\frac{\omega_1}{\omega_3} \right|
\end{equation}
and for $j=1,3$,
\begin{equation} \label{5.4}
Q_j(\zeta)=t_j Q(t_j\zeta+x_0),
\end{equation}
where $t_j=\omega_j/|\omega_j|$. Subsequently, the point $x_0$ will be
chosen in such a way that no pole of $Q_j$, $j=1,3$ lies on the real
axis. (This is equivalent to the requirement that no pole of $Q$
lies on
the line through the points $x_0$ and $x_0+ 2\omega_1$ nor on the line
through $x_0$ and $x_0+2\omega_3$. Since $Q$ has only finitely many
poles in the fundamental period parallelogram this can always be
achieved.) For such a choice of $x_0$ we infer that the entries of
$Q_j(\zeta)$ are real-analytic and periodic of period $\Omega_j
=2|\omega_j|$ whenever $\zeta$ is restricted to the real axis.
Using the
variable transformation $x=t_j\zeta+x_0$, $\psi(x)=\chi(\zeta)$ one
concludes that $\psi$ is a solution of
\begin{equation} \label{5.5}
J\psi'(x)+Q(x)\psi(x)=E\psi(x)
\end{equation}
if and only if $\chi$ is a solution of
\begin{equation} \label{5.6}
J\chi'(\zeta) + Q_j(\zeta)\chi(\zeta) = \lambda \chi(\zeta),
\end{equation}
where $\lambda=t_j E$.

Theorem \ref{t3.2} is now applicable and yields the following result.
\begin{thm}\label{t5.3}
Let $j=1$ or $3$. Then all $4\omega_j$-periodic (i.e., all
$2\omega_j$-periodic and all $2\omega_j$-semi-periodic) eigenvalues
associated with $Q$ lie in the strip $S_j$ given by
\begin{equation} \label{5.7}
S_j=\{E\in\bb C \, | \, |\Im(t_jE)|\le C_j \}
\end{equation}
for suitable constants $C_j>0$. The angle between the axes of the
strips $S_1$ and $S_3$ equals $\theta \in (0,\pi)$.
\end{thm}

Theorem \ref{t5.3} applies to any elliptic potential $Q$ whether or not
it is algebro-geometric. Next we present our principal result, a
characterization of all elliptic algebro-geometric potentials of the
AKNS hierarchy. Given the preparations in Sections
\ref{gauge}--\ref{fingap}, the proof of our principal result, Theorem
\ref{t5.4} below, will be fairly short.

\begin{thm}\label{t5.4}
$Q$ is an elliptic algebro-geometric AKNS potential if and only if
it is
a Picard-AKNS potential.
\end{thm}

\begin{proof}
The fact that any elliptic algebro-geometric AKNS potential is a Picard
potential is a special case of Theorem~\ref{6.3}.

Conversely, assume that $Q$ is a Picard-AKNS potential. Choose $R>0$
large enough such that the exterior of the closed disk
$\overline{D(0,R)}$ of radius $R$ centered at the origin contains no
intersection of $S_1$ and $S_3$ (defined in \eqref{5.7}), that is,
\begin{equation} \label{5.21}
(\bb C\backslash \overline{D(0,R)}) \cap(S_1\cap S_3)=\emptyset.
\end{equation}
Let $\rho_{j,\pm}(\lambda)$ be the Floquet multipliers of $Q_j$, that
is, the solutions of
\begin{equation} \label{5.22}
\rho^2_j-\tr (T_j) \rho_j+1=0.
\end{equation}
Then \eqref{5.21} implies that for $E\in\bb C
\backslash\overline{D(0,R)}$ at most one of the eigenvalues
$\rho_1(t_1E)$ and $\rho_3(t_3E)$ can be degenerate. In particular, at
least one of the operators $T_1$ and $T_3$ has distinct eigenvalues.
Since by hypothesis $Q$ is Picard, Picard's Theorem \ref{t5.1} applies
with $A=-J(Q-E)$ and guarantees the existence of two linearly
independent solutions $\psi_1(E,x)$ and $\psi_2(E,x)$ of $J\psi' +
Q\psi=E\psi$ which are elliptic of the second kind. Then
$\chi_{j,k}(\zeta)=\psi_k(t_j\zeta+x_0)$, $k=1,2$ are linearly
independent Floquet solutions associated with $Q_j$. Therefore the
points $\lambda$ for which $J \chi'+Q_j \chi=\lambda \chi$ has only one
Floquet solution are necessarily contained in $\overline{D(0,R)}$ and
hence finite in number. This is true for both $j=1$ and $j=3$. Applying
Theorem \ref{t4.2} then proves that both $Q_1$ and $Q_3$ are
algebro-geometric. This implies that $Q$ itself is algebro-geometric.
\end{proof}

The following corollary slightly extends the class of AKNS potentials
$Q(x)$ considered thus far in order to include some cases which
are not elliptic but very closely
related to elliptic $Q(x)$. Such cases have recently been considered
by Smirnov \cite{58g}.

\begin{cor}\label{c5.5}
Suppose
\begin{equation}
Q(x)=\begin{pmatrix} 0&-i q(x) \e^{-2(ax+b)}\\ip(x)\e^{2(ax+b)}&0
\end{pmatrix}
\end{equation}
where $a,b\in\bb C$ and $p,q$ are elliptic functions with a common
period lattice. Then $Q$ is an algebro-geometric AKNS potential if
and
only if $J\Psi'+Q\Psi=E\Psi$ has a meromorphic fundamental system of
solutions (with respect to the independent variable) for all values
of the spectral parameter $E\in\bb C$.
\end{cor}

\begin{proof}
Suppose that for all values of $E$ the equation $L\Psi=J\Psi'+Q\Psi
=E\Psi$ has a meromorphic fundamental system of solutions. Let
\begin{equation}\label{05051}
\mathcal T=
\begin{pmatrix}\e^{ax+b} & 0 \\ 0 & \e^{-ax+b} \end{pmatrix}.
\end{equation}
Then $\mathcal T L \mathcal T^{-1}=\tilde L+ia I=Jd/dx+\tilde Qia I$,
where
\begin{equation}
\tilde Q=\begin{pmatrix}0 &-iq\\ ip & 0\end{pmatrix}.
\end{equation}
Moreover, $L\Psi=E\Psi$ is equivalent to $\tilde L(\mathcal T\Psi)
=(E-ia)(\mathcal T\Psi)$. Hence the equation
$\tilde L\Psi=(E-ia)\Psi$
has a meromorphic fundamental system of solutions for all $E$.
Consequently, Theorem \ref{t5.4} applies and yields that
$\tilde Q$ is an algebro-geometric
AKNS potential. Thus, for some $n$ there exists a differential
expression $\tilde P$ of order $n+1$ with leading coefficient $-J^{n}$
such that $[\tilde P,\tilde L]=0$. Define $P=\mathcal T^{-1} \tilde
P\mathcal T$. The expression $P$ is a differential expression of order
$n+1$ with leading coefficient $-J^{n}$ and satisfies $[P,L]=\mathcal
T^{-1}[\tilde P,\tilde L+ia I]
\mathcal T=0$, that is, $Q$ is an algebro-geometric AKNS potential.
The
converse follows by reversing the above proof.
\end{proof}

We add a series of remarks further illustrating the significance of
Theorem \ref{t5.4}.

\begin{rem} \label{r5.5}
While an explicit proof of the algebro-geometric property of $(p,q)$ is
in general highly nontrivial (see, e.g., the references cited in
connection with special cases such as the Lam\'e-Ince and
Treibich-Verdier potentials in the introduction), the fact of whether or
not $J\Psi'(x)+Q(x)\Psi(x)= E\Psi(x)$ has a fundamental system of
solutions meromorphic in $x$ for all but finitely many values of the
spectral parameter $E\in\bb C$ can be decided by means of an elementary
Frobenius-type analysis (see, e.g., \cite{29} and \cite{30}). To date,
Theorem \ref{t5.4} appears to be the only effective tool to identify
general elliptic algebro-geometric solutions of the AKNS hierarchy.
\end{rem}

\begin{rem} \label{r5.6}
Theorem \ref{t5.4} complements Picard's Theorem \ref{t5.1} in the
special case where $A(x)=-J(Q(x)-E)$ in the sense that it determines
the
elliptic matrix functions $Q$ which satisfy the hypothesis of the
theorem precisely as (elliptic) algebro-geometric solutions of the
stationary AKNS hierarchy.
\end{rem}

\begin{rem} \label{r5.7}
Theorem \ref{t5.4} is also relevant in the context of the Weierstrass
theory of reduction of Abelian to elliptic integrals, a subject that
attracted considerable interest, see, for instance, \cite{5}, \cite{6},
\cite{6a}, Ch. 7, \cite{7}, \cite{8}, \cite{13d}, \cite{18}--\cite{20},
\cite{38}, \cite{40}, \cite{44}, \cite{44c}, \cite{58d}, \cite{58},
\cite{59}. In particular, the theta functions corresponding to the
hyperelliptic curves derived from the Burchnall-Chaundy polynomials
\eqref{2.22}, associated with Picard potentials, reduce to
one-dimensional theta functions.
\end{rem}

\section{Examples} \label{Ex}
\setcounter{equation}{0}
With the exception of the studies by Christiansen, Eilbeck, Enol'skii,
and Kostov in \cite{13d} and Smirnov in \cite{58g}, not too many
examples of elliptic solutions $(p,q)$ of the AKNS hierarchy
associated
with higher (arithmetic) genus curves of the type \eqref{2.22} have
been
worked out in detail. The genus $n=1$ case is considered, for example,
in \cite{37a}, \cite{51}. Moreover, examples for low genus $n$ for
special cases such as the nonlinear Schr\"odinger and mKdV equation
(see
\eqref{2.47} and \eqref{2.49}) are considered, for instance, in
\cite{4b}, \cite{6}, \cite{44d}, \cite{45a}, \cite{50}, \cite{58f}.
Subsequently we
will illustrate how the Frobenius method, whose essence is
captured by Proposition \ref{p552}, can be used to establish
existence of
meromorphic solutions and hence, by Theorem \ref{t5.4}, proves
their algebro-geometric property. The notation established in the
beginning of Section \ref{gauge} will be used freely in the following.

\begin{exmp} \label{e7.1}
Let
\begin{equation}
p(x)=q(x)=n(\zeta(x)-\zeta(x-\omega_2)-\eta_2),
\end{equation}
where $n\in\bb N$. The potential $(p,q)$ has two poles in the
fundamental period parallelogram. Consider first the pole $x=0$. In
this
case we have
\begin{equation}
R=\begin{pmatrix} 0&n\\n&0\end{pmatrix},
\end{equation}
whose eigenvalues are $\pm n$, that is, $\lambda=n$. Moreover,
since $p=q$ is
odd, we have $p_{2j-1}=q_{2j-1}=0$. One proves by induction that
$b_{2j}^{(2)}$ is a multiple of $(1,1)^t$ and that $b_{2j-1}^{(2)}$
is a
multiple of $(1,-1)^t$. Hence $b_{2n-1}^{(2)}$ is a multiple of
$(1,-1)^t$, that is, it is in the range of $R-n$. Hence every
solution of
$L\Psi=E\Psi$ is meromorphic at zero regardless of $E$.

Next consider the pole $x=\omega_2$ and shift coordinates by
introducing
$\xi=x-\omega_2$. Then we have $p(x)=q(x)
=n(\zeta(\xi+\omega_2)-\zeta(\xi)-\eta_2) =-p(\xi)$ and hence
\begin{equation}
R=\begin{pmatrix} 0&-n\\-n&0\end{pmatrix}.
\end{equation}
One can use again a proof by induction to show that $b_{2n-1}^{(2)}$
is in the range
of $R-n$, which is spanned by $(1,1)^t$.

Hence we have shown that the matrix
\begin{equation}
Q(x)=\begin{pmatrix}0&-in(\zeta(x)-\zeta(x-\omega_2)-\eta_2)\\
in(\zeta(x)-\zeta(x-\omega_2)-\eta_2)&0\end{pmatrix}
\end{equation}
is a Picard-AKNS and therefore an algebro-geometric AKNS potential.
\end{exmp}

\begin{exmp} \label{e7.2}
Here we let $p=1$ and $q=n(n+1)\wp(x)$, where $n\in\bb N$. Then
we have
just one pole in the fundamental period parallelogram. In this case we
obtain
\begin{equation}
R=\begin{pmatrix} 1&n(n+1)\\1&0\end{pmatrix}
\end{equation}
and $\lambda=n+1$. Since $q$ is even we infer that $q_{2j-1}=0$. A
proof by
induction then shows that $b_{2j}^{(2)}$ is a multiple of
$(n-2j,1)^t$
and that
$b_{2j-1}^{(2)}$ is a multiple of $(1,0)^t$. In particular,
$b_{2n}^{(2)}$ is a multiple of $(-n,1)^t$, which spans the range of
$R-\lambda$. This shows that
\begin{equation}
Q(x)=\begin{pmatrix}0&-in(n+1)\wp(x)\\
i&0\end{pmatrix}
\end{equation}
is a Picard-AKNS and hence an algebro-geometric AKNS potential.
\end{exmp}

Incidentally, if $p=1$, then $J\Psi'+Q\Psi=E\Psi$ is equivalent to the
scalar equation $\psi_2''-q\psi_2=-E^2\psi_2$ where
$\Psi=(\psi_1,\psi_2)^t$ and $\psi_1=\psi_2'-iE \psi_2$. Therefore, if
$-q$ is an elliptic algebro-geometric potential of the KdV hierarchy
then by Theorem 5.7 of \cite{32a} $\psi_2$ is meromorphic for all
values
of $E$. Hence $\Psi$ is meromorphic for all values of $E$ and therefore
$Q$ is a Picard-AKNS and hence an algebro-geometric AKNS potential.
Conversely if $Q$ is an algebro-geometric AKNS potential with $p=1$
then
$-q$ is an algebro-geometric potential of the KdV hierarchy (cf.
\eqref{2.51}). In particular, $q(x)=n(n+1)\wp(x)$ is the celebrated
class of Lam\'e potentials associated with the KdV hierarchy
(cf., e.g.,
\cite{29} and the references therein).

\begin{exmp} \label{e7.3}
Suppose $e_2=0$ and hence $g_2=4e_1^2$ and $g_3=0$. Let
$u(x)=-\wp'(x)/(2e_1)$. Then, near $x=0$,
\begin{equation}
u(x)=\frac{1}{e_1 x^3}-\frac{e_1}{5}x +O(x^3),
\end{equation}
and near $x=\pm\omega_2$,
\begin{equation}
u(x)=e_1(x\mp\omega_2)-\frac{3e_1^3}{5}(x\mp\omega_2)^5
+O((x\mp\omega_2)^7).
\end{equation}

Now let $p(x)=3u(x)$ and $q(x)=u(x-\omega_2).$ Then $p$ has a
third-order pole at $0$ and a simple zero at $\omega_2$ while $q$ has a
simple zero at zero and a third-order pole at $\omega_2$. Let us first
consider the point $x=0$. We have
\begin{equation}
R=\begin{pmatrix} -2&e_1\\3/e_1&0\end{pmatrix},
\end{equation}
and hence $\lambda=1$. Moreover, $p_2=q_2=0$, $p_4=-3e_1/5$, and
$q_4=-3e_1^3/5$. Since $\lambda=1$ we have to show that $b_3^{(2)}$ is a
multiple of $(q_0,-1)^t$. We get, using $p_2=q_2=0$,
\begin{equation}
b_3^{(2)}=(q_0 E^4/6+q_4,-E^4/6-q_0p_4)^t,
\end{equation}
which is a multiple of $(q_0,-1)^t$ if and only if $q_4=p_4q_0^2$, a
relationship which is indeed satisfied in our example.

Next consider the point $x=\omega_2$. Changing variables to
$\xi=x-\omega_2$ and using the periodic properties of $u$ we find that
$p(x)=3q(\xi)$ and $q(x)=p(\xi)/3$. Thus $q$ has a pole at $\xi=0$
and one obtains $m=2$, $p_0=3e_1$, $q_0=1/e_1$, $p_2=q_2=0$,
$p_4=-9e_1^3/5$,
and $q_4=-e_1/5$. Since $\lambda=3$, we have to compute again
$b_3^{(2)}$
and find, using $p_2=q_2=0$,
\begin{equation}
b_3^{(2)}=(-q_0 E^4/6+3q_4,-E^4/2-q_0p_4)^t,
\end{equation}
which is a multiple of $(q_0,-3)^t$ if and only if $9q_4=p_4q_0^2$,
precisely what we need.

Hence, if $e_2=0$ and $u(x)=-\wp'(x)/(2e_1)$, then
\begin{equation}
Q(x)=\begin{pmatrix}0&-iu(x-\omega_2)\\3iu(x)&0\end{pmatrix}
\end{equation}
is a Picard-AKNS and therefore an algebro-geometric AKNS potential.
\end{exmp}

\begin{exmp} \label{e7.4}
Again let $e_2=0$. Define $p(x)=\frac23(\wp''(x)-e_1^2)$
and $q(x)=-\wp(x-\omega_2)/e_1^2$. First consider $x=0$. We have
$m=-3$, $p_0=4$, $q_0=1$, $p_2=q_2=0$, $p_4=-2e_1^2/5$, and
$q_4=-e_1^2/5$. This yields $\lambda=1$ and we need to show that
$b_4^{(2)}$ is a multiple of $(1,-1)^t$.
We find, using $p_2=q_2=0$ and $q_0=\lambda=1$,
\begin{equation}
b_4^{(2)}=i(-E^5/24-p_4/4-q_4,E^5/24+5p_4/4-q_4)^t.
\end{equation}
This is a multiple of $(1,-1)^t$ if $2q_4=p_4$, which is indeed
satisfied.

Next consider $x=\omega_2$. Now $q$ has a second-order pole, that
is, we
have $m=1$. Moreover,
\begin{equation}
q(x)=\frac{-1}{e_1^2}(\frac{1}{(x-\omega_2)^2}+\frac{e_1^2}{5}
(x-\omega_2)^2+ O((x-\omega_2)^2)
\end{equation}
and
\begin{equation}
p(x)=-2e_1^2+ 96 e_1^4 (x-\omega_2)^4+ O((x-\omega_2)^6).
\end{equation}
We now need $b_2^{(2)}$ to be a multiple of $(q_0,-2)^t$, which is
satisfied for $q_2=p_2=0$.

Hence, if $e_2=0$, then
\begin{equation}
Q(x)=\begin{pmatrix}0&i\wp(x-\omega_2)/e_1^2\\
2i(\wp''(x)-e_1^2)/3&0\end{pmatrix}
\end{equation}
is a Picard-AKNS and thus an algebro-geometric AKNS potential.
\end{exmp}


\end{document}